\documentclass[%
 reprint,
nofootinbib,
 amsmath,amssymb,
 aps,
]{revtex4-2}
\usepackage{fontspec}
\usepackage[english]{babel}
\babelprovide[import,onchar=ids fonts]{japanese}
\babelfont[japanese]{rm}[
  BoldFont=HaranoAjiMincho-Bold.otf,
  ItalicFont=HaranoAjiMincho-Regular.otf,
  BoldItalicFont=HaranoAjiMincho-Bold.otf
]{HaranoAjiMincho-Regular.otf}
\babelfont[japanese]{sf}[
  BoldFont=HaranoAjiGothic-Bold.otf,
  ItalicFont=HaranoAjiGothic-Medium.otf,
  BoldItalicFont=HaranoAjiGothic-Bold.otf
]{HaranoAjiGothic-Medium.otf}

\usepackage{graphicx}
\usepackage{dcolumn}
\usepackage{bm}
\usepackage[unicode,hidelinks]{hyperref}

\usepackage{orcidlink}
\usepackage{comment}

\usepackage{physics}

\usepackage[normalem]{ulem} 

\usepackage{aas_macros}

\usepackage{booktabs} 
\usepackage{siunitx}  
\usepackage{enumitem}

\usepackage{multirow} 

\usepackage{tikz}
\usepackage{xcolor}
\definecolor{RED}{rgb}{1,0,0} 

\usepackage{ragged2e} 

\newcommand{\rei}[1]{}

\newcommand{\Rei}[1]{}

\newcommand{\rEi}[1]{}

\begin{document}
\title{Fast Radio Bursts from Induced Upscattering. I. Magnetar Polar Fireballs}

\author{Rei Nishiura\orcidlink{0009-0003-8209-5030}}
 \email{rei.nishiura@yukawa.kyoto-u.ac.jp}
 \affiliation{%
  Center for Gravitational Physics and Quantum Information, 
 Yukawa Institute for Theoretical Physics, Kyoto University, Kyoto 606-8502, Japan}%
 \author{Kunihito Ioka\orcidlink{0000-0002-3517-1956}}%
 \email{kunihito.ioka@yukawa.kyoto-u.ac.jp}
\affiliation{%
 Center for Gravitational Physics and Quantum Information, 
 Yukawa Institute for Theoretical Physics, Kyoto University, Kyoto 606-8502, Japan}%
 \author{Bing Zhang\orcidlink{0000-0002-9725-2524}}%
\email{bzhang1@hku.hk}
\affiliation{%
The Hong Kong Institute for Astronomy and Astrophysics, The University of Hong Kong,
Pokfulam Road, Hong Kong}
\affiliation{%
Department of Physics, The University of Hong Kong,
Pokfulam Road, Hong Kong}%

%
%
\date{\today}

\begin{abstract}
The extreme brightness temperatures of fast radio bursts (FRBs) demand
a coherent process that converts magnetar burst energy into
millisecond radio emission. We propose that FRBs can be produced
within magnetar magnetospheres by induced (stimulated) upscattering
in a relativistic fireball outflow. In this paper, we focus on induced
Compton upscattering in an electron--positron fireball launched from
the polar region of a magnetar. Radiation from the X-ray burst
accelerates the polar fireball outflow to Lorentz factors of
$\Gamma\sim10$--$10^3$. Low-frequency seed waves, such as fast
magnetosonic or Alfvén waves, entering the outflow from the side or
against its motion undergo induced scattering and are Lorentz-boosted
by the relativistic bulk motion from $\nu_0\sim10\mathrm{kHz}$ to
$\nu_{\rm obs}\sim\Gamma^2\nu_0\sim\mathrm{GHz}$. Combining the
polar fireball dynamics with the linear growth and nonlinear
saturation of the induced scattering instability, we derive the FRB
luminosity $L_{\rm FRB}^{\rm iso}$ and fractional spectral bandwidth $\Delta\nu/\nu_{\rm c}$. By considering the
possible range of pair densities and temperatures, we find that this
polar channel can account for the luminosity of Galactic FRB~20200428
and reach luminosities comparable to those of the brightest
extragalactic FRBs. If the pair-loaded region is narrower than the
relativistic beaming angle $1/\Gamma$, the model also produces a
narrow spectrum and predicts $L_{\rm FRB}^{\rm
iso}\propto(\Delta\nu/\nu_{\rm c})^2$, a trend
consistent with observations of repeating FRBs. The induced
upscattering model thus connects magnetar X-ray bursts to FRBs
through a coherent emission process grounded in plasma physics and
makes observationally testable predictions.
\end{abstract}

\maketitle

\section{Introduction}
\label{sec:introduction}
Fast radio bursts (FRBs) are millisecond-duration flashes and the most luminous radio transients in the Universe \citep{Lorimer2007-frb,Thornton2013-frb}. Their brightness temperatures can reach $T_{\rm b}\sim10^{35}\,{\rm K}$, far above the limit of any incoherent radiation process. The radiation must therefore be produced coherently, yet its physical mechanism remains one of the central unsolved problems in high energy astrophysics. FRB dispersion measures also provide cosmological probes of ionized matter, as anticipated before their discovery and now demonstrated with localized bursts \citep{Ioka2003-dm,Inoue2004-dm,Macquart2020-baryons}.

The discovery in 2020 of FRB~20200428 from the Galactic magnetar SGR~1935+2154, simultaneous with a hard X-ray burst, demonstrated that magnetars can produce FRB-like radio bursts and established them as the leading candidate sources of at least a substantial fraction of FRBs \citep{Andersen2020-iz,Bochenek2020-ev,Mereghetti2020-ka,Li2021-nd,Ridnaia2021-xray,Tavani2021-xray,2021NatAs...5..414K}. Magnetar activity is powered by magnetic energy and may be triggered by crustal failure, magnetic reconnection, or a coupled disturbance of the crust and magnetosphere \citep{1995MNRAS.275..255T,2020ApJ...897....1L,Yuan2022-gk}. Because the magnetic energy density increases rapidly toward the neutron-star surface, the primary dissipation is expected to occur in the inner magnetosphere. The released energy can initially appear as thermal radiation and an optically thick fireball containing abundant electron and positron ($e^\pm$) pairs \citep{2001ApJ...561..980T,Ioka2020-pn}, as a Poynting-flux-dominated pulse \citep{2020ApJ...897....1L,Yuan2020-tu,Yuan2022-gk,2020MNRAS.494.2385K}, or as a combination of these components. It must then be transported outward by radiation, plasma kinetic energy, and Poynting flux. Surface motions and magnetospheric disturbances naturally excite low-frequency waves with characteristic frequencies up to $\sim10\,{\rm kHz}$ \citep{Blaes1989-starquakes}; thus, GHz radiation is likely produced only after the released energy has propagated away from the stellar surface.

Growing observational evidence places at least some FRB emission within, or not far beyond, the magnetosphere \citep{2020MNRAS.498.1397L} (see for a review \citep{2023RvMP...95c5005Z}). Microsecond and even submicrosecond temporal structure implies an extremely compact emitting region \citep{Nimmo2021-microstructure,Majid2021-nanostructure,Snelders2023-ultrafast}. Narrow spectra with $\Delta\nu/\nu_{\rm c}\ll 1$ have been observed in repeating FRBs \citep{2022RAA....22l4001Z,2023ApJ...955..142Z} and have been argued to support a compact magnetospheric origin \citep{2024ApJ...974..160K}. Diverse polarization properties, including pulsar-like position-angle swings, orthogonal position-angle jumps, and strong circular polarization, have also been observed \citep{Luo2020-pa,Mckinven2024-pa,2024ApJ...972L..20N,2024NSRev..12E.293J}, with theoretical studies exploring their magnetospheric origins \citep{2023MNRAS.522.2448Q,2026ApJ...997...37Q}. Scintillation of FRB~20221022A further constrains its transverse source size \citep{2025Natur.637...48N} and disfavors emission at very large radii \citep{1992ApJ...390..454H,2014MNRAS.442L...9L,2016MNRAS.461.1498M,2017ApJ...842...34W,2020MNRAS.494.4627M,2019MNRAS.485.4091M,2020ApJ...896..142B,2021MNRAS.500.2704Y,2024PhRvL.132c5201I}. A magnetospheric model must nevertheless explain how a GHz wave escapes through the dense, strongly magnetized plasma to radii where its dimensionless strength parameter $a_0=eE/(m_{\mathrm{e}} c\omega_0)$ falls below unity. Strong waves may undergo nonlinear steepening, mode conversion, or intense particle scattering and damping \citep{Lyubarsky2003-fms,2020ApJ...897....1L,Beloborodov2022-ultrastrong,Chen2022-strongfms,2023ApJ...959...34B,2024ApJ...975..223B,2023ApJ...957..102G}. Their propagation and induced-scattering opacity therefore depend sensitively on the plasma outflow, the wave--field angle, and nonlinear saturation, and remain active theoretical issues \citep{Qu2022-transparency,Sobacchi2024-escape,kmdy-17md,2026PhRvD.114f3022I}.

A promising class of magnetospheric models is based on inverse-Compton-like frequency conversion, in which a low-frequency electromagnetic (EM) or magnetohydrodynamics (MHD) wave is scattered by relativistic plasma and its kinetic energy is converted into coherent high-frequency radiation \citep{Zhang2022-kh,2023MNRAS.522.2448Q}. Related realizations have also been discussed in the language of free-electron lasers \citep{Lyutikov2020-fel,Lyutikov2021-fel}. A burst can readily excite waves with $\nu_0\sim10\,{\rm kHz}$ throughout the magnetosphere, and a fast magnetosonic (FMS) wave can propagate across the background magnetic field and enter an outflow. Scattering by plasma with bulk Lorentz factor $\Gamma$ then boosts the frequency by two Lorentz transformations,
\begin{equation}
\label{eq:seed_wave_require}
\nu_{\rm FRB}\sim \Gamma^2\nu_0
\simeq 1\,{\rm GHz}\,
\Gamma_{2.5}^{2}
\left(\frac{\nu_0}{10\,{\rm kHz}}\right),
\end{equation}
where $\Gamma_{2.5}\equiv\Gamma/10^{2.5}$. The enormous observed brightness requires the scattering process to be coherent and hence induced (stimulated), rather than spontaneous single-particle scattering. Hereafter, we use the term ``induced scattering.''

Induced scattering in the strong magnetic fields relevant to magnetar magnetospheres has only recently been formulated quantitatively for $e^\pm$ pair plasmas \citep{2025PhRvD.111f3055N,Nishiura2026-nx} (see also \citep{2008ApJ...682.1443L}). Physically, it is a three-wave parametric interaction among an incident pump wave, a scattered EM wave, and a plasma density mode. The beat of the two EM waves exerts a ponderomotive force that drives the density perturbation, which in turn scatters the pump and closes the feedback loop. When the frequency and wavevector matching conditions are satisfied, the occupation of the scattered mode stimulates further scattering and produces exponential growth. The growth rate can be calculated from the plasma properties. In a strongly magnetized $e^\pm$ plasma, the relevant branches include induced Compton, Brillouin \citep{2024PhRvE.110a5205I}, and Raman scattering, among which induced Compton scattering can dominate in magnetar magnetospheres \citep{kmdy-17md}. Three wave interactions of force-free MHD waves may also be important \citep{1998PhRvD..57.3219T,2019ApJ...881...13L,2019MNRAS.483.1731L,2023ApJ...957..102G}, but induced Compton scattering may be more relevant to our scenario, as discussed in Sec. \ref{sec:seed-waves-competing-processes}. The analytic linear growth rates have also been confirmed by particle-in-cell (PIC) simulations \citep{tvyv-yn1z,kmdy-17md}. As the instability evolves, the resonant interaction flattens the plasma distribution function, causing the growth to saturate before complete scattering. This nonlinear saturation must therefore be included when connecting the plasma physics to an FRB \citep{tvyv-yn1z,kmdy-17md,2026PhRvD.114f3022I}.

In this paper, we construct an FRB model grounded in this plasma physics. Within the inverse-Compton-like framework, we apply the recently derived growth rate of induced scattering in a strongly magnetized $e^\pm$ pair plasma and follow its nonlinear saturation. We refer to the resulting mechanism as the \emph{induced upscattering model}: low-frequency waves already present in the magnetosphere are amplified and shifted to GHz frequencies by a relativistic outflow.

This work is the first in a series exploring two principal channels by which energy released near the neutron-star surface can accelerate the required plasma. Here we study the polar channel, as illustrated in Fig. \ref{fig:fireball_outflow}. Near a magnetic pole, the field lines are approximately radial, so X-rays emerging from the base of a pair fireball exert a sustained outward radiation force. Cyclotron-resonant and Thomson scattering can accelerate the $e^\pm$ outflow to $\Gamma\sim10$--$10^3$ \citep{2023MNRAS.519.4094W,Wada2025-ve}. The resulting relativistic plasma then upscatters low-frequency magnetospheric waves through induced scattering. We focus on a pure pair fireball in this paper and leave baryon-loaded fireballs for future work.

The complementary channel operates away from the magnetic poles. Because the field lines there are not radial, radiation from the stellar surface cannot accelerate the plasma to a sufficiently large Lorentz factor and may even decelerate it \citep{2002ApJ...574..332T}. An Alfv\'en wave launched at the surface can instead grow nonlinear as it propagates outward, reaching $\delta B/B\gtrsim1$ and ejecting a plasmoid \citep{Yuan2020-tu,Yuan2022-gk}. Magnetic pressure can then accelerate the plasmoid \citep{2010PhRvE..82e6305L,2010ApJ...720.1490L,2011MNRAS.411.1323G,2023MNRAS.519..497T} until it can upscatter ambient low frequency waves through induced scattering. We will investigate this off-polar plasmoid fireball channel in a subsequent paper.\footnote{While preparing this paper, we became aware of independent recent work \citep{Beloborodov2026-narrow}.}

For the polar pair-fireball channel, we derive the FRB luminosity and fractional bandwidth $\Delta\nu/\nu_{\rm c}$ and compare them with observations. In particular, if the accelerated, pair-loaded plasma is confined to an angular region narrower than the relativistic beaming angle $1/\Gamma$, the model naturally produces a narrow spectrum, $\Delta\nu/\nu_{\rm c}<1$, and predicts
\begin{equation}
L_{\rm FRB}^{\rm iso}\propto
\left(\frac{\Delta\nu}{\nu_{\rm c}}\right)^2.
\end{equation}
As shown in Sec. \ref{sec:small}, this trend is consistent with the observed luminosity--bandwidth distributions within current uncertainties \citep{2022MNRAS.515.3577H,2023MNRAS.526.3652K,2024MNRAS.52710425S}. At fixed X-ray burst luminosity, increasing the pair load raises the FRB luminosity. As shown in Fig. \ref{fig:fireball_outflow_luminosity}, the maximum luminosity allowed by the model is comparable to the largest FRB luminosities observed to date \citep{Shah_2026}.

The remainder of this paper is organized as follows. Section~II summarizes the induced upscattering scenario. Section~III develops the pair-fireball dynamics, including photon decoupling, pair freeze out, and radiative acceleration of the polar outflow. Section~IV applies induced Compton scattering to the seed wave, evaluates its linear growth and nonlinear saturation, and derives the FRB luminosity. Section~V studies how a localized pair-loading region controls the spectral bandwidth and its correlation with luminosity. We discuss the implications and limitations of the model in Sec.~VI and summarize our conclusions in Sec.~VII.

\section{Model Overview}
\label{sec:model-overview}

\begin{figure*}
\centering
\includegraphics[width=\textwidth]{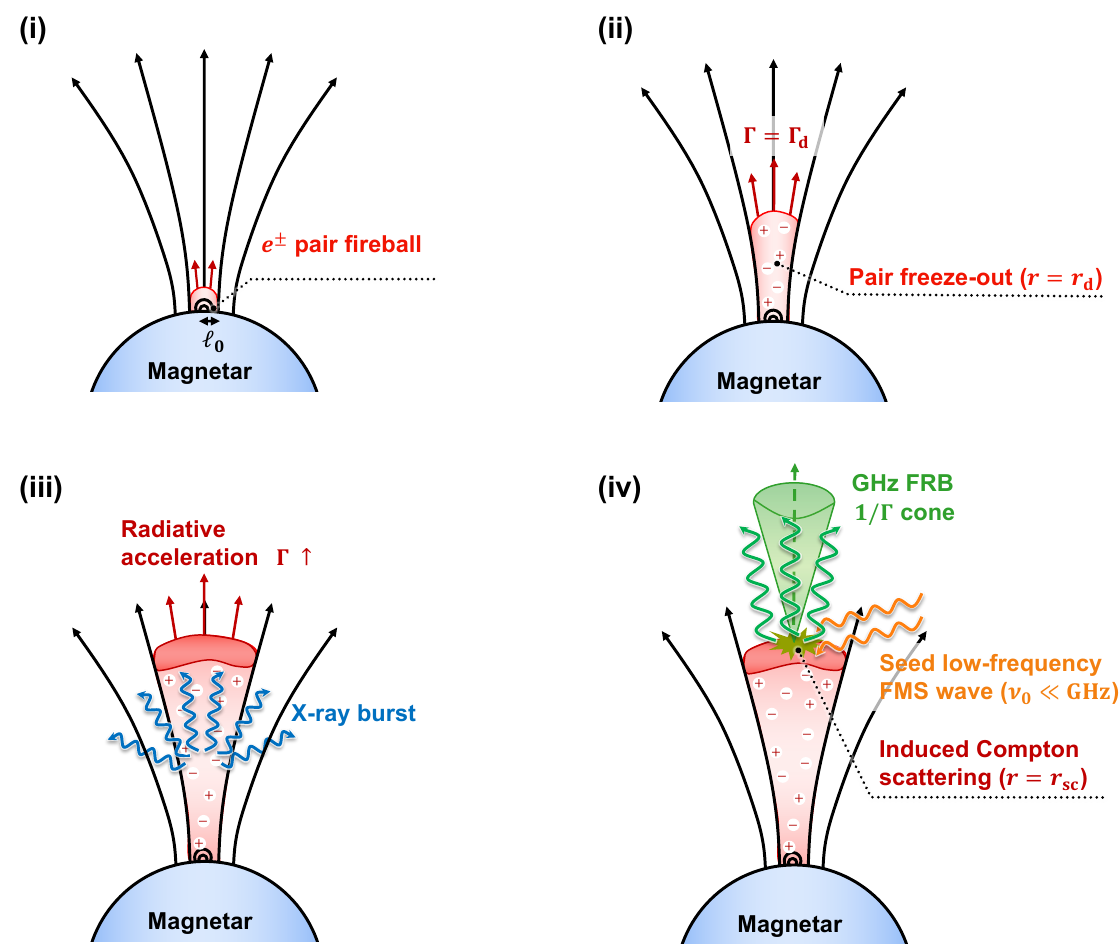}
\caption{\justifying
Schematic illustration of the FRB emission scenario considered in this work, in which induced Compton upscattering operates in a polar fireball outflow.
(i) An optically thick $e^\pm$ fireball forms in the polar region of the magnetar surface and expands along a dipolar magnetic flux tube.
(ii) As the fireball expands, photons decouple from the $e^\pm$ plasma, and the pair density freezes out near the decoupling radius as pair annihilation becomes inefficient.
(iii) The escaping X-ray radiation accelerates the $e^\pm$ plasma to ultrarelativistic velocities through radiative acceleration, particularly through cyclotron resonant scattering and Thomson scattering.
(iv) FMS denotes a fast magnetosonic wave. A low frequency seed FMS wave incident on the outflow undergoes induced Compton scattering in the comoving frame. If the seed wave is incident from the side or against the bulk motion, the scattered wave is shifted to the GHz band through Lorentz transformations and can be observed as an FRB.
}
\label{fig:fireball_outflow}
\end{figure*}

We consider an FRB emission scenario in which induced Compton scattering operates in a polar fireball outflow through the four stages illustrated in Fig.~\ref{fig:fireball_outflow}. First, a high energy burst in the polar region of the magnetar surface produces abundant electron and positron pairs, forming an optically thick $e^\pm$ fireball. The fireball subsequently expands along a dipolar magnetic flux tube. The formation and initial expansion of the fireball are discussed in detail in Sec.~\ref{sec:initial-pair-fireball}.

As the fireball expands and its optical depth decreases, photons begin to decouple from the $e^\pm$ plasma. In a strong magnetic field, the scattering cross section of X-ray photons depends on their polarization mode, and different modes therefore have different diffusion times \citep{2023MNRAS.519.4094W}. Photons begin to escape efficiently when the comoving diffusion time becomes smaller than the comoving dynamical time. The pair freeze out radius is also comparable to the photon decoupling radius. Near this radius, the pair annihilation time becomes comparable to the comoving dynamical time, and pair annihilation becomes inefficient at larger radii. Photon decoupling and pair freeze out are discussed in Sec.~\ref{sec:photon-decoupling-pair-freezeout}.

After photon decoupling, the escaping X-ray radiation further accelerates the $e^\pm$ plasma. The X-ray photons approach the plasma predominantly from behind and exert an outward radiation force. In particular, radiative acceleration through cyclotron resonant scattering and Thomson scattering can accelerate the plasma to a terminal Lorentz factor of $\Gamma_{\infty}\sim10$--$10^3$ \citep{2023MNRAS.519.4094W,Wada2025-ve}. We consider this ultrarelativistic fireball outflow as the scattering medium for induced Compton scattering. The density evolution after freeze out, radiative acceleration, and kinetic luminosity of the outflow are discussed in Sec.~\ref{sec:post-freezeout-kinetic-luminosity}.

We finally consider low frequency seed waves incident on the ultrarelativistic fireball outflow. We assume that kHz--MHz MHD waves can be supplied by magnetic reconnection, magnetic disturbances, or a cascade of MHD turbulence in the magnetar magnetosphere. The energy required for such a seed wave is much smaller than the characteristic energy budgets associated with the background magnetic field and the X-ray burst. The frequency, amplitude, and energy requirements of the seed wave are discussed in Sec.~\ref{sec:seed-wave-properties}.

When the seed wave is incident from the side or against the bulk motion, relativistic aberration makes the wave appear nearly head-on in the comoving frame. The seed wave then undergoes induced Compton scattering with the $e^\pm$ plasma. The scattered wave is amplified while retaining approximately the seed wave frequency in the comoving frame. In the lab frame, however, the Lorentz transformations into and out of the comoving frame produce a large frequency boost, allowing a low frequency seed wave to emerge as a GHz FRB. We refer to this induced Compton scattering process, in which the bulk motion produces the frequency increase, as \emph{induced Compton upscattering}. The linear growth, nonlinear saturation, and resulting FRB luminosity are discussed in Sec.~\ref{sec:induced-compton-upscattering}.

In the next section, we evaluate the physical quantities associated with each stage shown in Fig.~\ref{fig:fireball_outflow} using order of magnitude estimates.

\section{Pair Fireball and Polar Outflow Dynamics}
\label{sec:pair-fireball-polar-outflow}
\subsection{Initial Pair Fireball}
\label{sec:initial-pair-fireball}

We first examine whether the X-ray burst associated with Galactic FRB 20200428 can plausibly produce an optically thick pair fireball near the magnetar surface \citep{Ioka2020-pn,2023MNRAS.519.4094W}. We adopt the Galactic FRB 20200428 as our fiducial case and explicitly retain the relevant parameter dependences in the following discussion. We model the X-ray emitting region near the magnetic pole as a hemispherical blackbody with a characteristic radius $\ell_0$. The burst reached a peak luminosity of $\sim 10^{41}~\mathrm{erg~s^{-1}}$ and had a characteristic cutoff energy of $\sim 80~\mathrm{keV}$ \citep{Li2021-nd} (see Eq. \eqref{eq:Compton_observation}). For an order of magnitude estimate, we approximate the radiation by a blackbody with a temperature $T_0$. The radiative flux is then given by
\begin{equation}
\label{eq:xray-flux}
\begin{aligned}
F_{\mathrm{X}}
&= \sigma_{\rm SB} T_0^4
= \frac{L_{\mathrm{X}}}{2\pi \ell_0^2}.
\end{aligned}
\end{equation}
The characteristic radius of the emitting region is therefore expressed as
\begin{equation}
\label{eq:initial-footpoint-radius}
\begin{aligned}
\ell_0
= \left(
\frac{L_{\mathrm{X}}}
{2\pi \sigma_{\rm SB}T_0^4}
\right)^{\frac{1}{2}}\sim 1.7\times10^4\,{\rm cm}
\frac{L_{\mathrm{X},41}^{\frac{1}{2}}}{T_{0,9}^{2}}.
\end{aligned}
\end{equation}

The compact size of the emitting region implies a very large radiative compactness. The radiation energy density near the source is $u_{\mathrm{X}}=L_{\mathrm{X}}/(2\pi \ell_0^2c)$. We define an effective photon number density in units of the electron rest mass energy as
\begin{equation}
\label{eq:effective-photon-density}
\begin{aligned}
n_{\gamma0}
\equiv \frac{u_{\mathrm{X}}}{m_{\mathrm{e}} c^2}= \frac{L_{\mathrm{X}}}
{2\pi \ell_0^2 m_{\mathrm{e}} c^3}.
\end{aligned}
\end{equation}
The corresponding compactness parameter, which characterizes the importance of interactions over a length scale $\ell_0$, is given, using Eqs.~\eqref{eq:initial-footpoint-radius} and \eqref{eq:effective-photon-density} by
\begin{equation}
\label{eq:initial-compactness}
\begin{aligned}
\mathcal{C}_0
&\equiv n_{\gamma 0}\sigma_{\mathrm{T}}\ell_0= \frac{\sigma_{\mathrm{T}} L_{\mathrm{X}}}
{2\pi \ell_0 m_{\mathrm{e}} c^3}
\\
&\sim 1\times10^7
\frac{L_{\mathrm{X},41}}{\ell_{0,4}}
\gg 1.
\end{aligned}
\end{equation}
The large value of $\mathcal{C}_0$ implies that only a small fraction of the radiative energy needs to be converted into photons capable of producing pairs for pair loading to become dynamically important. In the following, we assume that an optically thick pair fireball forms near the magnetar surface \citep{1995MNRAS.275..255T,Ioka2020-pn,2023MNRAS.519.4094W}.

The subsequent expansion of the optically thick pair fireball is controlled by the geometry of the polar magnetic flux tube \citep{1986ApJ...308L..43P,2001ApJ...561..980T}. The radial distance normalized by the neutron star radius $R$ is defined as
\begin{equation}
\label{eq:normalized-radius}
x \equiv \frac{r}{R}.
\end{equation}
We also define the dimensionless comoving temperature of the plasma as
\begin{equation}
\label{eq:dimensionless-temperature}
\Theta
\equiv \frac{k_B T'(r)}{m_{\mathrm{e}} c^2}.
\end{equation}
While the background magnetic field remains approximately dipolar, its radial dependence is described by
\begin{equation}
B = B_{\mathrm{p}} x^{-3}.
\label{eq:dipole_magnetic}
\end{equation}
Combining Eq.~\eqref{eq:dipole_magnetic} with magnetic flux conservation gives the radial evolution of the flux tube radius as
\begin{equation}
\label{eq:flux-tube-radius}
\ell = \ell_0 x^{\frac{3}{2}}.
\end{equation}

For a radiation dominated outflow confined to the dipolar flux tube, conservation of the energy and entropy fluxes determines the acceleration and comoving temperature of the fireball \citep{1986ApJ...308L..43P,2001ApJ...561..980T}. The Lorentz factor is given by
\begin{equation}
\label{eq:fireball-lorentz-factor}
\Gamma = x^{\frac{3}{2}}.
\end{equation}
The corresponding comoving temperature is expressed as
\begin{equation}
\label{eq:fireball-temperature}
\Theta = \Theta_0 x^{-\frac{3}{2}}.
\end{equation}
The Lorentz factor and the comoving temperature therefore have opposite radial dependences. The decrease in the comoving temperature is compensated by the increase in the bulk Lorentz factor. The characteristic observed temperature consequently scales as $T_{\mathrm{obs}}\propto \Gamma T'=\mathrm{const}$. Thus, during this acceleration regime, adiabatic expansion does not reduce the characteristic photon energy measured by a distant observer. 

Finally, the density of the $e^\pm$ plasma is determined by pair equilibrium while the fireball remains sufficiently optically thick. When the pairs are nonrelativistic in the comoving frame and predominantly occupy the lowest Landau level, the total comoving pair number density follows from the Landau degeneracy per unit area, $eB/(2\pi\hbar c)$, together with the phase space density along the background magnetic field. The resulting Boltzmann distribution is described by \citep{2023MNRAS.519.4094W}
\begin{equation}
\label{eq:pair-equilibrium-density}
\begin{aligned}
n_\pm'
= \frac{e B m_{\mathrm{e}}}
{(2\pi^3)^{\frac{1}{2}}\hbar^2}
\Theta^{\frac{1}{2}}
\exp\!\left(-\frac{1}{\Theta}\right).
\end{aligned}
\end{equation}
Here, the total number density of electrons and positrons is defined as
\begin{equation}
\label{eq:total-pair-density}
n_\pm = n_+ + n_-.
\end{equation}

\subsection{Photon Decoupling and Pair Freeze-out}
\label{sec:photon-decoupling-pair-freezeout}

As the pair fireball expands along the dipolar magnetic flux tube, photons eventually decouple from the $e^\pm$ plasma. We estimate the decoupling radius $r_{\mathrm{d}}$ by comparing the comoving diffusion time with the comoving dynamical time. In the regime where the cyclotron frequency 
\begin{equation}
\label{eq:cyclotron-frequency-decoupling}
\omega_{\mathrm{c}}\equiv\frac{eB}{m_{\mathrm{e}}c},
\end{equation}
is much larger than the characteristic X-ray frequency in the comoving frame, the scattering cross section of X-mode photons are strongly suppressed and therefore decouple first. 
We adopt the Rosseland mean scattering cross section for the X-mode in the form \citep{1992herm.book.....M,2002MNRAS.332..199L}
\begin{equation}
\label{eq:xmode-rosseland-cross-section}
\sigma_{\mathrm{X}}
= \frac{4\pi^2}{5}\sigma_{\mathrm{T}}
\left(\Theta\frac{B_{\mathrm{Q}}}{B}\right)^2,
\end{equation}
where,
\begin{equation}
\label{eq:qed-critical-field}
\begin{aligned}
B_{\mathrm{Q}}
= \frac{m_{\mathrm{e}}^2 c^3}{e\hbar}= 4.414\times10^{13}\,{\rm G},
\end{aligned}
\end{equation}
is the quantum critical magnetic field.

We estimate the photon escape time by considering diffusion across the transverse size $\ell$ of the fireball. The comoving diffusion time is given by \citep{Ioka2020-pn,2023MNRAS.519.4094W}
\begin{equation}
\label{eq:photon-diffusion-time}
t_{\rm diff}'
\sim n_\pm'\sigma_{\mathrm{X}}\frac{\ell^2}{c}.
\end{equation}
The comoving dynamical time is expressed as
\begin{equation}
\label{eq:dynamical-time-decoupling}
t_{\rm dyn}'
\sim \frac{r}{c\Gamma}.
\end{equation}
We define the ratio of these two time scales as
\begin{equation}
\label{eq:diffusion-dynamical-ratio}
\mathcal{D}\equiv \frac{t_{\rm diff}'}{t_{\rm dyn}'}.
\end{equation}
Using the radial scalings in Eqs. \eqref{eq:dipole_magnetic},\eqref{eq:flux-tube-radius},  \eqref{eq:fireball-lorentz-factor}, and  \eqref{eq:fireball-temperature} to Eqs. \eqref{eq:pair-equilibrium-density}, \eqref{eq:xmode-rosseland-cross-section}, \eqref{eq:photon-diffusion-time}, and \eqref{eq:dynamical-time-decoupling}, we identify the onset of photon decoupling by the condition $\mathcal{D}=1$. The corresponding radius and Lorentz factor can be estimated as
\begin{equation}
\label{eq:decoupling-radius-lorentz-factor}
\begin{aligned}
\left\{
\begin{aligned}
\Gamma_{\mathrm{d}}
&\equiv \Gamma(r_{\mathrm{d}})\sim 3.0,\\
r_{\mathrm{d}}
&=2.1\times10^6\,{\rm cm}.
\end{aligned}
\right.
\end{aligned}
\end{equation}

Assuming that pair creation becomes inefficient after photon decoupling, pair annihilation continues to reduce the pair density until its time scale becomes comparable to the expansion time. The comoving number density of positrons formally evolves according to
\begin{equation}
\label{eq:positron-annihilation-evolution}
\frac{\dd n_+'}{\dd t}
\sim -n_+'n_-'\sigma_{\rm ann}v_{\rm rel}.
\end{equation}
The same argument applies to electrons. Here, $v_{\rm rel}$ denotes the mean relative velocity between electrons and positrons. In the nonrelativistic limit, the annihilation cross section is given by (e.g. \citep{2011hea..book.....L})
\begin{equation}
\label{eq:annihilation-cross-section}
\sigma_{\rm ann}
\sim \frac{3}{8}\frac{c}{v_{\rm rel}}
\sigma_{\mathrm{T}}.
\end{equation}
Using Eqs. \eqref{eq:total-pair-density}, \eqref{eq:positron-annihilation-evolution}, and \eqref{eq:annihilation-cross-section}, the annihilation time in the comoving frame is expressed as
\begin{equation}
\label{eq:annihilation-time}
\begin{aligned}
t_{\rm ann}'
\equiv \frac{n_+'}
{\left\lvert \dd n_+' / \dd t\right\rvert}\sim \frac{16}{3}
\frac{1}{n_\pm'\sigma_{\mathrm{T}} c}.
\end{aligned}
\end{equation}

We estimate the pair density at freeze out. We further assume that the pair freeze out radius is comparable to the photon decoupling radius, $r_{\mathrm{fo}}\sim r_{\mathrm{d}}$. Under this approximation, the pair density at $r_{\mathrm{d}}$ is obtained by equating the annihilation and dynamical timescales given by Eqs. \eqref{eq:annihilation-time} and \eqref{eq:dynamical-time-decoupling},\footnote{With our notation, $\Gamma_{\mathrm{d},0.5}=1$ corresponds to $\Gamma_{\mathrm{d}}=10^{0.5}\simeq 3.16$. Based on the estimate in Eq. \eqref{eq:decoupling-radius-lorentz-factor}, we adopt $\Gamma_{\mathrm{d}}=3$ in the following order-of-magnitude estimates. This small difference does not affect any of our conclusions.}
\begin{equation}
\label{eq:pair-density-decoupling}
n_{\pm,\mathrm{d}}'
\sim \frac{16}{3}
\frac{\Gamma_{\mathrm{d}}^{\frac{1}{3}}}
{\sigma_{\mathrm{T}}R}
\sim 1.2\times10^{19}\,{\rm cm}^{-3}
\frac{\Gamma_{\mathrm{d},0.5}^{\frac{1}{3}}}{R_6}.
\end{equation}
Here, we used the acceleration law from Eq. \eqref{eq:fireball-lorentz-factor},
\begin{equation}
\label{eq:gamma-decoupling-scaling}
\Gamma_{\mathrm{d}}
\sim \left(\frac{r_{\mathrm{d}}}{R}\right)^{\frac{3}{2}}.
\end{equation}
The comoving temperature at the decoupling radius is then expressed from Eq. \eqref{eq:fireball-temperature} as
\begin{equation}
\label{eq:temperature-decoupling}
\begin{aligned}
\Theta_{\mathrm{d}}
&= \Theta_0
\left(\frac{r_{\mathrm{d}}}{R}\right)^{-\frac{3}{2}}
\\
&= 5.6\times10^{-2}
\frac{T_{0,9}}
{\Gamma_{\mathrm{d},0.5}}.
\end{aligned}
\end{equation}

\subsection{Post Freeze-out and Kinetic Luminosity}
\label{sec:post-freezeout-kinetic-luminosity}

After pair freeze out, conservation of the total lepton number flux determines the radial evolution of the pair density. We assume that both pair creation and annihilation become negligible beyond the decoupling radius $r_{\mathrm{d}}$. The pair density at $r>r_{\mathrm{d}}$ can then be obtained from particle number conservation. This density is particularly important because it determines the energy carried by the outflow and hence the energy budget available for conversion into FRB emission by induced scattering. For a relativistic outflow with $\beta\simeq1$, conservation of the total lepton number flux between $r_{\mathrm{d}}$ and $r>r_{\mathrm{d}}$ is expressed as
\begin{equation}
\label{eq:post-freeze-particle-flux}
\pi \ell(r)^2 n_\pm'(r)\Gamma(r)
= \pi \ell_{\mathrm{d}}^2
n_{\pm,\mathrm{d}}'\Gamma_{\mathrm{d}}.
\end{equation}
The radial evolution of the flux tube radius follows from Eq.~\eqref{eq:flux-tube-radius} and is given by
\begin{equation}
\label{eq:post-freeze-flux-tube-radius}
\ell
= \ell_{\mathrm{d}}
\left(\frac{r}{r_{\mathrm{d}}}\right)^{\frac{3}{2}}.
\end{equation}
Combining Eqs.~\eqref{eq:post-freeze-particle-flux} and \eqref{eq:post-freeze-flux-tube-radius} with the pair density at the decoupling radius gives the comoving number density of electrons and positrons at radius $r$ as
\begin{equation}
\label{eq:standard-pair-density}
\begin{aligned}
n_{\pm,{\rm con}}'(r)
&\sim \frac{16}{3}
\frac{\Gamma_{\mathrm{d}}^{\frac{10}{3}}R^2}
{\sigma_{\mathrm{T}}\Gamma r^3}
\\
&\sim 3.3\times10^{14}\,{\rm cm}^{-3}
\frac{\Gamma_{\mathrm{d},0.5}^{\frac{10}{3}}R_6^2}
{r_8^3\Gamma}.
\end{aligned}
\end{equation}
We adopt Eq.~\eqref{eq:standard-pair-density} as the conservative density of the pair-fireball case. We also define the density enhancement factor as
\begin{equation}
\label{eq:density-enhancement}
\xi
\equiv \frac{n_\pm'}{n_{\pm,{\rm con}}'}.
\end{equation}
Here, $\xi$ is a phenomenological enhancement relative to the post freeze-out density in Eq.~\eqref{eq:standard-pair-density}. Additional pair loading after freeze out, for example through magnetic reconnection, can lead to $\xi\gg1$ and thereby increase the energy carried by the outflow. Its attainable value depends on the location and rate of pair injection, as discussed in Sec.~\ref{sec:plasma-supply-thermal-evolution}.

After photon decoupling, the $e^\pm$ plasma can be further accelerated by the X-ray radiation escaping from the fireball. In particular, cyclotron resonant scattering and Thomson scattering can accelerate the plasma to a terminal Lorentz factor of $\Gamma_{\infty}\sim10$--$10^3$ \citep{2023MNRAS.519.4094W,Wada2025-ve}. We treat $\Gamma_{\infty}$ as a parameter below. We estimate the radial evolution of the Lorentz factor by assuming that the plasma approximately follows the equilibrium Lorentz factor at which the net radiation force vanishes in the comoving frame. We assume that the X-ray radiation emerging at $r=r_{\mathrm{d}}$ has an opening angle of $\sim1/\Gamma_{\mathrm{d}}$. The corresponding transverse size is estimated as
\begin{equation}
\label{eq:transverse-size-decoupling}
r_{\mathrm{d}}\theta_{\mathrm{d}}
\sim \frac{r_{\mathrm{d}}}{\Gamma_{\mathrm{d}}}.
\end{equation}
As viewed from a distance $r\gg r_{\mathrm{d}}$, the radiation source has an apparent angular size given by
\begin{equation}
\label{eq:opening-angle}
\begin{aligned}
\tan\theta_{\mathrm{s}}
\sim \theta_{\mathrm{s}}\sim \frac{1}{r-r_{\mathrm{d}}}
\frac{r_{\mathrm{d}}}{\Gamma_{\mathrm{d}}}\sim \frac{r_{\mathrm{d}}}
{r\Gamma_{\mathrm{d}}}.
\end{aligned}
\end{equation}
The equilibrium Lorentz factor at which the radiation force approximately vanishes in the comoving frame is therefore estimated from Eq.~\eqref{eq:opening-angle} as
\begin{equation}
\label{eq:post-freeze-lorentz-factor}
\begin{aligned}
\Gamma(r)
\sim \frac{1}{\theta_{\mathrm{s}}}\sim \Gamma_{\mathrm{d}}\frac{r}{r_{\mathrm{d}}}.
\end{aligned}
\end{equation}
For a plasma element moving at an angle $\theta>\theta_{\mathrm{s}}$ relative to the central direction of the radiation field, Compton drag limits its Lorentz factor to $\Gamma\sim1/\sin\theta$. In contrast, for $\theta<\theta_{\mathrm{s}}$, radiative acceleration can maintain the plasma near the equilibrium Lorentz factor in Eq.~\eqref{eq:post-freeze-lorentz-factor}.  Assuming that the plasma follows Eq.~\eqref{eq:post-freeze-lorentz-factor} until it reaches $\Gamma_{\infty}$, the corresponding coasting radius is defined as
\begin{equation}
\label{eq:acceleration-radius}
\begin{aligned}
r_{\rm acc}
= \frac{\Gamma_\infty}{\Gamma_{\mathrm{d}}}
 r_{\mathrm{d}}\sim 2.2\times10^8\,{\rm cm}
\frac{R_6\Gamma_{\infty,2.5}}
{\Gamma_{\mathrm{d},0.5}^{\frac{1}{3}}}.
\end{aligned}
\end{equation}

At $r\gtrsim r_{\rm acc}$, where the Lorentz factor reaches $\Gamma_{\infty}$, the isotropic kinetic luminosity is estimated using Eqs.~\eqref{eq:standard-pair-density}, \eqref{eq:density-enhancement}, and \eqref{eq:post-freeze-lorentz-factor} as
\begin{equation}
\label{eq:kinetic-luminosity-iso}
\begin{aligned}
L_{\rm kin}^{\rm iso}(r)
&\sim 4\pi r^2\Gamma^2 n_\pm'(r)m_{\mathrm{e}}c^3
\\
&=3.2\times10^{42}\,{\rm erg}\,{\rm s}^{-1}
\frac{\Gamma_{\infty,2.5}
\Gamma_{\mathrm{d},0.5}^{\frac{10}{3}}\xi_5R_6^2}{r_9}.
\end{aligned}
\end{equation}
As discussed in Sec.~\ref{sec:plasma-supply-thermal-evolution}, the corresponding kinetic energy requirement is well below the kinetic energy available in observed energetic magnetar giant flares. If the region accelerated to $\Gamma_{\infty}$ is confined within an opening angle of $\sim1/\Gamma_{\infty}$, the corresponding true kinetic luminosity is estimated from the solid angle correction as\footnote{ Strictly speaking, since the plasma moves along dipolar flux tubes, its position and velocity directions do not coincide. This geometrical effect is treated explicitly in Sec.~\ref{sec:small}. Since it changes the true luminosity by less than a factor of $2$, we neglect it here. }
\begin{equation}
\label{eq:kinetic-luminosity-true}
\begin{aligned}
L_{\rm kin}^{\rm true}(r)
\sim \frac{L_{\rm kin}^{\rm iso}(r)}{4\Gamma_{\infty}^2}=7.9\times10^{36}\,{\rm erg}\,{\rm s}^{-1}
\frac{\Gamma_{\mathrm{d},0.5}^{\frac{10}{3}}R_6^2\xi_5}
{\Gamma_{\infty,2.5}r_9}.
\end{aligned}
\end{equation}

\section{Induced Compton upscattering}
\label{sec:induced-compton-upscattering}
\subsection{Seed Wave Properties}
\label{sec:seed-wave-properties}

As illustrated in Fig.~\ref{fig:fireball_outflow}(iv), we consider a scenario in which a seed wave at low frequency enters the fireball outflow and undergoes induced Compton scattering in the comoving frame. The scattering is nearly elastic in the comoving frame, whereas the Lorentz transformations into and out of this frame can substantially increase the wave frequency measured in the lab frame. The scattered wave can therefore emerge in the GHz band and be observed as an FRB. We assume that the seed wave propagates as a linear eigenmode of the strongly magnetized $e^\pm$ plasma. In the low frequency limit $\omega_0'\ll\omega_{\mathrm{c}}$, the relevant transverse modes are the Alfv\'en and FMS waves. For propagation parallel to the background magnetic field, the Alfv\'en wave is purely transverse and can be represented by either linear or circular polarization. The FMS wave can propagate obliquely to the background magnetic field and is linearly polarized, with its electric field perpendicular to the background magnetic field.

We consider an FMS wave as the incident seed wave in the following analysis. A large frequency boost requires the seed wave not to co-propagate with the outflow. Thus, a strictly outward, field-aligned wave does not provide the required geometry, although a field-aligned Alfvén wave can contribute if it propagates inward. In the low-frequency limit considered here, the Alfvén and FMS branches are degenerate for propagation parallel to the background magnetic field in a strongly magnetized $e^\pm$ plasma.

We adopt a fiducial observing frequency of $1.4\,\mathrm{GHz}$ and estimate the corresponding seed wave frequency required in our scenario. For a fireball outflow moving with a bulk Lorentz factor $\Gamma$, the frequency of an incident FMS seed wave in the comoving frame is given by
\begin{equation}
\label{eq:seed-frequency-transform}
\nu_0'
= \Gamma(1-\beta\cos\theta_0)\nu_0\sim\Gamma\nu_0.
\end{equation}
Here, $\theta_0$ denotes the angle between the wavevector of the seed wave and the bulk velocity in the lab frame. We assume that the bulk velocity is approximately parallel to the local background magnetic field. In the strongly magnetized $e^\pm$ plasma, induced Compton scattering changes the wave frequency only slightly in the comoving frame. The frequency of the scattered wave is expressed as
\begin{equation}
\label{eq:scattered-frequency-comoving}
\begin{aligned}
\nu_1'
\sim \nu_0'\left(1-2\Theta^{\frac{1}{2}}\right)\sim \nu_0'.
\end{aligned}
\end{equation}
where we assume a nonrelativistic plasma comoving temperature, $\Theta\ll1$. If $\theta_1'$ denotes the angle between the scattered wavevector and the bulk velocity measured in the comoving frame, its observed frequency is given from Eqs. \eqref{eq:seed-frequency-transform} and \eqref{eq:scattered-frequency-comoving} by
\begin{equation}
\label{eq:observed-frequency-transform}
\begin{aligned}
\nu_{\rm obs}
= \Gamma(1+\beta\cos\theta_1')\nu_1'\sim \Gamma^2\nu_0.
\end{aligned}
\end{equation}
The last relation applies to the geometry of interest. Thus, for a given observed frequency $\nu_{\rm obs}$, the required seed wave frequency is estimated as
\begin{equation}
\label{eq:seed-frequency}
\begin{aligned}
\nu_0
\sim 14\,{\rm kHz}
\left(\frac{\nu_{\rm obs}}{1.4\,{\rm GHz}}\right)\frac{1}{\Gamma_{\infty,2.5}^2}.
\end{aligned}
\end{equation}
For later use, the corresponding angular frequency of the seed wave in the comoving frame is expressed as
\begin{equation}
\label{eq:seed-angular-frequency-comoving}
\begin{aligned}
\omega_0'
=2\pi\nu_0'=2.8\times10^7\,{\rm s}^{-1}
\left(\frac{\nu_{\rm obs}}{1.4\,{\rm GHz}}\right)
\frac{1}{\Gamma_{\infty,2.5}}.
\end{aligned}
\end{equation}

The linear growth rate and nonlinear evolution of induced Compton scattering adopted in this work are applicable when the initial relative amplitude of the seed wave 
\begin{equation}
\label{eq:seed-amplitude}
\eta_{\rm seed}'
\equiv \frac{\delta B_{\rm seed}'}{B_{\mathrm{d}}},
\end{equation}
is smaller than unity \citep{2025PhRvD.111f3055N,Nishiura2026-nx},
\begin{equation}
\label{eq:seed-amplitude_linearity}
\eta_{\rm seed}'<1.
\end{equation}
We therefore restrict our analysis to this regime. Under this condition, we show below that the isotropic luminosity of the seed wave can remain much smaller than the magnetic and X-ray energy. Nevertheless, even such a weak seed wave can produce the large isotropic luminosities of bright extragalactic FRBs through induced Compton upscattering, as shown in Eq.~\eqref{eq:frb-kinetic-relation}. Possible mechanisms that generate or supply the seed wave are discussed in Sec.~\ref{sec:seed-waves-competing-processes}. 

We show that the condition in Eq.~\eqref{eq:seed-amplitude_linearity} corresponds to a modest energy requirement for the seed wave. The local dipole magnetic field is expressed as
\begin{equation}
\label{eq:dipole-field-radius}
\begin{aligned}
B_{\mathrm{d}}(r)
= B_{\mathrm{p}}\frac{R^3}{r^3}\sim 2\times10^5\,{\rm G}
\frac{B_{\mathrm{p},14.3}R_6^3}{r_9^3}.
\end{aligned}
\end{equation}
We assume that the fireball moves approximately along the local dipolar magnetic field. Since the Lorentz boost is then parallel to the background magnetic field, its magnetic field strength is unchanged between the two frames, $B_{\mathrm{d}}'=B_{\mathrm{d}}$. For the seed wave, the Lorentz transformation gives
\begin{equation}
\delta B_{\rm seed}'=\Gamma(1-\beta\cos\theta_0)\delta B_{\rm seed}\sim\Gamma\delta B_{\rm seed}.
\end{equation}
Hence, $\eta_{\rm seed}'\sim\Gamma\eta_{\rm seed}$, where $\eta_{\rm seed}\equiv\delta B_{\rm seed}/B_{\mathrm{d}}$. The linearity condition in Eq.~\eqref{eq:seed-amplitude_linearity} therefore requires $\eta_{\rm seed}<1/\Gamma$ in the lab frame.

The corresponding isotropic luminosity of the seed wave can be estimated from its local energy flux, using Eq.~\eqref{eq:seed-amplitude}, 
\begin{equation}
\label{eq:maximum_seed_wave}
\begin{aligned}
L_{\mathrm{seed}}^{\mathrm{iso}}&\sim4\pi r^2\frac{c\delta B_{\rm seed}^2}{4\pi}\sim cr^2B_{\mathrm{d}}^2\frac{\eta_{\rm seed}'^2}{\Gamma^2}\\
&<\frac{cr^2B_{\mathrm{d}}^2}{\Gamma^2}=1.2\times10^{34}~\mathrm{erg~s}^{-1}\frac{B_{\mathrm{p},14.3}^2R_6^6}{r_9^4\Gamma_{\infty,2.5}^2},
\end{aligned}
\end{equation}
in the region where $\Gamma\simeq\Gamma_\infty$. For comparison, we define a reference isotropic Poynting luminosity corresponding to a magnetic perturbation with an amplitude comparable to the local dipole field,
\begin{equation}
L_{\mathrm{mag}}^{\mathrm{iso}}\sim cr^2B_{\mathrm{d}}^2\sim1.2\times10^{39}~\mathrm{erg~s}^{-1}\frac{B_{\mathrm{p},14.3}^2R_6^6}{r_9^4}.
\end{equation}
The peak X-ray luminosity is also much larger, $\sim10^{41}\,\mathrm{erg\,s^{-1}}$. Thus, even a seed wave near the upper end of the linear regime in Eq.~\eqref{eq:seed-amplitude_linearity} carries only a small fraction of either $L_{\mathrm{mag}}^{\mathrm{iso}}$ and $L_\mathrm{X}$. The linear seed wave assumed here therefore requires only a modest energy budget compared with the magnetic and X-ray energy available in the magnetosphere. In the following, we assume that such an FMS seed wave is present.

\subsection{Linear Growth and the Scattering Onset Radius}
\label{sec:linear-growth-scattering-onset}

Induced scattering initially amplifies the scattered wave through exponential linear growth and subsequently enters a nonlinear stage in which the growth saturates, as demonstrated by PIC simulations \citep{tvyv-yn1z,kmdy-17md}. The simulations indicate that the transition to nonlinear evolution occurs after approximately 10 e-foldings of the scattered wave. In this section, we use the linear growth rate to estimate the scattering onset radius $r_{\mathrm{sc}}$, which is defined below in Eq.~\eqref{eq:scattering-onset-radius}. At this radius, a seed FMS wave incident on the fireball outflow is sufficiently amplified by induced scattering to enter nonlinear evolution in the comoving frame.

In a strongly magnetized $e^\pm$ plasma, induced Compton scattering can occur through neutral and charged modes \citep{2025PhRvD.111f3055N,Nishiura2026-nx}. The neutral mode describes a density fluctuation without charge separation, whereas the charged mode involves charge separation between electrons and positrons. The neutral mode can dominate for the parameter range mainly considered here, and we therefore focus on this mode below. We discuss the relative importance of the two modes in Sec.~\ref{sec:seed-waves-competing-processes}.  

For a broadband seed wave, the maximum linear growth rate of induced Compton scattering through the neutral mode is \citep[Eq.~(119)]{2025PhRvD.111f3055N}
\begin{equation}
\label{eq:neutral-growth-rate}
\gamma_{\mathrm{N}}'
\sim \pi
\frac{\eta_{\rm seed}'^{\,2}\omega_0'}{\sigma_B'}.
\end{equation}
Here, $\sigma_B'$ is the magnetization parameter defined in Eq.~\eqref{eq:magnetization-parameter}. For simplicity, we adopt $\Delta\omega_0'\sim\omega_0'$.
More generally, the broadband growth rate contains the factor
$(\omega_0'/\Delta\omega_0')^2$ \citep{2025PhRvD.111f3055N}.
A narrower incident spectrum therefore enhances the growth and
moves the scattering onset inward. We neglect this additional dependence in our
order-of-magnitude estimates. We next evaluate the physical quantities required to estimate the e-folding number obtained by normalizing the linear growth rate by the comoving dynamical time.

The plasma frequency of the $e^\pm$ plasma in the fireball outflow is obtained from Eqs.~\eqref{eq:standard-pair-density} and \eqref{eq:density-enhancement} and is expressed as
\begin{equation}
\label{eq:plasma-frequency}
\begin{aligned}
\omega_{\mathrm{p}}'
\equiv
\left(\frac{4\pi e^2n_\pm'}{m_{\mathrm{e}}}\right)^{\frac{1}{2}}
=5.7\times10^{11}\,{\rm s}^{-1}
\frac{\Gamma_{\mathrm{d},0.5}^{\frac{5}{3}}
R_6\xi^{\frac{1}{2}}_5}
{r_9^{\frac{3}{2}}\Gamma_{\infty,2.5}^{\frac{1}{2}}}.
\end{aligned}
\end{equation}
The cyclotron frequency is given by
\begin{equation}
\label{eq:cyclotron-frequency-radius}
\begin{aligned}
\omega_{\mathrm{c}}(r)
= \frac{eB_{\mathrm{d}}(r)}{m_{\mathrm{e}}c}
=3.5\times10^{12}\,{\rm s}^{-1}
\frac{B_{\mathrm{p},14.3}R_6^3}{r_9^3}.
\end{aligned}
\end{equation}
Since the Lorentz boost is parallel to the background magnetic field, Eq.~\eqref{eq:cyclotron-frequency-radius} is unchanged in the comoving frame. The magnetization parameter in the comoving frame is therefore obtained from Eqs.~\eqref{eq:plasma-frequency} and \eqref{eq:cyclotron-frequency-radius} as
\begin{equation}
\label{eq:magnetization-parameter}
\begin{aligned}
\sigma_B'
\equiv \left(\frac{\omega_{\mathrm{c}}}{\omega_{\mathrm{p}}'}\right)^2
=38~
\frac{B_{\mathrm{p},14.3}^2R_6^4\Gamma_{\infty,2.5}}
{r_9^3\Gamma_{\mathrm{d},0.5}^{\frac{10}{3}}\xi_5}.
\end{aligned}
\end{equation}

These quantities determine the amplification efficiency of induced Compton scattering. Substituting Eqs.~\eqref{eq:seed-angular-frequency-comoving} and \eqref{eq:magnetization-parameter} into Eq.~\eqref{eq:neutral-growth-rate}, we define the e-folding number accumulated over the comoving dynamical time as
\begin{equation}
\label{eq:neutral-efoldings}
\begin{aligned}
N_{\mathrm{N}}
\equiv \gamma_{\mathrm{N}}'t_{\rm dyn}'
&\sim 61~
\left(\frac{\eta_{\rm seed}'}{0.5}\right)^2
\left(\frac{\nu_{\rm obs}}{1.4\,{\rm GHz}}\right)
\\
&\quad\times
\frac{
r_9^4
\Gamma_{\mathrm{d},0.5}^{\frac{10}{3}}
\xi_5
}{
\Gamma_{\infty,2.5}^3
B_{\mathrm{p},14.3}^2
R_6^4
}.
\end{aligned}
\end{equation}
A value of $N_{\mathrm{N}}\sim1$ provides an approximate criterion for appreciable linear amplification within one dynamical time. PIC simulations indicate that the scattered wave enters the nonlinear evolution after approximately 10 e-foldings and subsequently approaches saturation \citep{tvyv-yn1z,kmdy-17md}. We therefore adopt
\begin{equation}
\label{eq:nonlinear_evolution}
N_{\mathrm{N}}=10,
\end{equation}
as the criterion for the onset of nonlinear evolution.

We define $r_{\mathrm{sc}}$ as the scattering onset radius at which $N_{\mathrm{N}}=10$. Using Eq.~\eqref{eq:neutral-efoldings}, this radius is estimated as
\begin{equation}
\label{eq:scattering-onset-radius}
\begin{aligned}
r_{\rm sc}
&=6.4\times10^8\,{\rm cm}
\left(\frac{\eta_{\rm seed}'}{0.5}\right)^{-\frac{1}{2}}
\left(\frac{\nu_{\rm obs}}{1.4\,{\rm GHz}}\right)^{-\frac{1}{4}}
\\
&\quad\times
\frac{\Gamma_{\infty,2.5}^{\frac{3}{4}}
B_{\mathrm{p},14.3}^{\frac{1}{2}}R_6}
{\Gamma_{\mathrm{d},0.5}^{\frac{5}{6}}\xi^{\frac{1}{4}}_5}.
\end{aligned}
\end{equation}
In the next section, we evaluate the kinetic luminosity of the fireball outflow at the scattering onset radius and estimate the FRB luminosity produced by induced Compton scattering.

\subsection{Nonlinear Saturation and FRB Luminosity}
\label{sec:nonlinear-saturation-frb-luminosity}

In this section, we show that, for the parameter range considered in our scenario, the scattered wave reaches nonlinear saturation before the seed wave is substantially attenuated. This corresponds to the partial scattering regime \citep{tvyv-yn1z,kmdy-17md}. In this regime, the FRB luminosity is primarily determined by the kinetic luminosity of the fireball outflow at the scattering onset radius $r_{\rm sc}$. Specifically, we derive
\begin{equation}
\label{eq:frb-kinetic-relation}
L_{\rm FRB}^{\rm iso}
\sim \Theta(r_{\rm sc})^{\frac{1}{2}}
L_{\rm kin}^{\rm iso}(r_{\rm sc}),
\end{equation}
when the seed wave energy density exceeds the saturation energy density in Eq.~\eqref{eq:saturation-energy-density}. Thus, the kinetic luminosity of the fireball outflow sets the characteristic FRB luminosity.

The nonlinear saturation of induced Compton scattering is controlled by the formation of a plateau in the particle distribution around the resonant velocity \citep{tvyv-yn1z,kmdy-17md}. As the scattered wave grows, energy is transferred from the incident wave to the plasma and modifies the particle distribution along the background magnetic field. The growth stops once a plateau develops around the resonant velocity. The energy density required to form this plateau is expected to be of the same order as the initial parallel internal energy density, 
\begin{equation}
\label{eq:thermal-energy-density}
u_{\rm th}'\equiv\frac{1}{2}n_\pm'k_BT'=\frac{1}{2}n_\pm'm_{\mathrm{e}}c^2\Theta.
\end{equation}
 This saturation mechanism has been confirmed by PIC simulations \citep{tvyv-yn1z,kmdy-17md}. The simulations followed the evolution of the scattered wave induced by an incident Alfv\'en (or FMS) wave propagating along the background magnetic field in a one dimensional periodic system. The system entered the nonlinear evolution after approximately $N_{\rm N}\sim10$ e-foldings of linear growth. For this one dimensional scattering geometry, the saturation energy density is estimated, using Eq. \eqref{eq:thermal-energy-density}, as
\begin{equation}
\label{eq:saturation-energy-density}
\begin{aligned}
u_{\rm sat}'
&\sim u_{\rm th}'
\times \frac{1}{2}\frac{v_{\mathrm{A}}'}{c}
\left(\frac{m_{\mathrm{e}}c^2}{k_BT'}\right)^{\frac{1}{2}}
\\
&\sim \frac{1}{4}n_\pm'm_{\mathrm{e}}c^2
\Theta^{\frac{1}{2}}
\end{aligned}
\end{equation}
where the comoving Alfv\'en velocity is defined by
\begin{equation}
\label{eq:alfven-velocity}
v_{\mathrm{A}}'
\equiv
\frac{c}
{\sqrt{1+\frac{\omega_{\mathrm{p}}'^2}{\omega_{\mathrm{c}}^2}}}.
\end{equation}

Two distinct nonlinear evolutions of induced Compton scattering have been identified, depending on the relative magnitudes of the saturation energy density and the seed (incident) wave energy density \citep{tvyv-yn1z,kmdy-17md}. If $u_{\rm seed}'\ge u_{\rm sat}'$, the scattered wave grows to $u_{\rm sat}'$ through energy transfer from the seed wave and saturates while part of the seed wave remains. If $u_{\rm seed}'<u_{\rm sat}'$, the seed wave is substantially attenuated before the scattered wave reaches the saturation energy density. These regimes are referred to as partial scattering and full scattering, respectively, and are summarized as
\begin{equation}
\label{eq:scattering-regimes}
\left\{
\begin{aligned}
u_{\rm seed}' &\ge u_{\rm sat}'
&&\Rightarrow {\rm partial\ scattering},\\
u_{\rm seed}' &< u_{\rm sat}'
&&\Rightarrow {\rm full\ scattering}.
\end{aligned}
\right.
\end{equation}

In our scenario, partial scattering is typically the relevant regime because the seed wave energy density is sufficiently large to reach saturation before the seed wave is substantially attenuated. To see this, the energy density of the seed wave in the comoving frame is obtained from Eq.~\eqref{eq:seed-amplitude} as
\begin{equation}
\label{eq:seed-energy-density}
u_{\rm seed}'
= \frac{B_{\mathrm{d}}(r)^2}{4\pi}
\eta_{\rm seed}'^{\,2}.
\end{equation}
The boundary between partial and full scattering is determined by $u_{\rm seed}'=u_{\rm sat}'$. Using Eqs.~\eqref{eq:saturation-energy-density} and \eqref{eq:seed-energy-density}, the corresponding seed wave amplitude is
\begin{equation}
\label{eq:saturation-seed-amplitude}
\begin{aligned}
\eta_{\rm seed}^{{\rm sat}\,'}(r_{\mathrm{sc}})
&= \left(
\frac{\pi n_\pm'm_{\mathrm{e}}c^2
\Theta^{\frac{1}{2}}}{B_{\mathrm{d}}^2}
\right)^{\frac{1}{2}}
\\
&\sim 1.3\times10^{-2}
\left(\frac{\eta_{\rm seed}'}{0.5}\right)^{-\frac{3}{4}}
\left(\frac{\nu_{\rm obs}}{1.4\,{\rm GHz}}\right)^{-\frac{3}{8}}
\\
&\quad\times
\frac{
\Gamma_{\mathrm{d},0.5}^{\frac{5}{12}}
\xi_5^{\frac{1}{8}}
\Theta_{-2}^{\frac{1}{4}}
\Gamma_{\infty,2.5}^{\frac{5}{8}}
}{
B_{\mathrm{p},14.3}^{\frac{1}{4}}
R_6^{\frac{1}{2}}
}.
\end{aligned}
\end{equation}
For the fiducial parameters, $\eta_{\rm seed}^{{\rm sat}\,'}\sim10^{-2}$ is much smaller than the fiducial seed amplitude $\eta_{\rm seed}'=0.5$, which remains consistent with the upper limit in Eq.~\eqref{eq:maximum_seed_wave}. The fiducial case therefore lies well within the partial scattering regime. The full scattering regime is discussed in Appendix~\ref{app:full-scattering}.

In the partial scattering regime, the scattered wave grows until its energy density reaches $u_{\rm sat}'$. Using Eq.~\eqref{eq:saturation-energy-density}, the corresponding energy flux in the lab frame is estimated as
\begin{equation}
\label{eq:saturation-flux}
\begin{aligned}
F_{\rm sat}
&= cu_{\rm sat}
\\
&\sim c\Gamma^2
\left\langle(1+\beta\cos\theta_1')^2\right\rangle_E
u_{\rm sat}'
\\
&\sim \Gamma^2n_\pm'm_{\mathrm{e}}c^3
\Theta^{\frac{1}{2}}.
\end{aligned}
\end{equation}
Here, $\langle\cdots\rangle_E$ denotes an angular average weighted by the scattered wave energy distribution $\dd u_{\rm sat}'/\dd\Omega'$. Assuming an isotropic distribution in the comoving frame, $\dd u_{\rm sat}'/\dd\Omega'=\mathrm{const}$, we combine this angular average with the numerical factor in Eq.~\eqref{eq:saturation-energy-density} and set the resulting order unity factor to unity. The FRB isotropic luminosity is then
\begin{equation*}
L_{\rm FRB}^{\rm iso}
\sim 4\pi r_{\rm sc}^2 F_{\rm sat}
\sim \Theta(r_{\rm sc})^{\frac{1}{2}}
L_{\rm kin}^{\rm iso}(r_{\rm sc}),
\end{equation*}
which recovers Eq.~\eqref{eq:frb-kinetic-relation}. For the fiducial seed amplitude $\eta_{\rm seed}'=0.5$, which lies well within the partial scattering regime, the kinetic luminosity at the scattering onset radius is obtained from Eqs.~\eqref{eq:scattering-onset-radius} and \eqref{eq:standard-pair-density} as
\begin{equation}
\label{eq:kinetic-luminosity-scattering}
\begin{aligned}
L_{\rm kin}^{\rm iso}(r_{\rm sc})
&=4\pi r_{\rm sc}^2\Gamma_\infty^2
n_\pm'(r_{\rm sc})m_{\mathrm{e}}c^3
\\
&=4.9\times10^{42}\,{\rm erg}\,{\rm s}^{-1}
\left(\frac{\eta_{\rm seed}'}{0.5}\right)^{\frac{1}{2}}
\\
&\quad\times
\left(\frac{\nu_{\rm obs}}{1.4\,{\rm GHz}}\right)^{\frac{1}{4}}
\frac{\Gamma_{\mathrm{d},0.5}^{\frac{25}{6}}
\Gamma_{\infty,2.5}^{\frac{1}{4}}R_6\xi_5^{\frac{5}{4}}}
{B_{\mathrm{p},14.3}^{\frac{1}{2}}}.
\end{aligned}
\end{equation}
For a fixed comoving plasma temperature of $\Theta=10^{-2}$, Eq.~\eqref{eq:frb-kinetic-relation} therefore gives
\begin{equation}
\label{eq:frb-luminosity-saturated}
\begin{aligned}
L_{\rm FRB}^{\rm iso}
&=\Theta^{\frac{1}{2}}L_{\rm kin}^{\rm iso}
\\
&=4.9\times10^{41}\,{\rm erg}\,{\rm s}^{-1}
\left(\frac{\eta_{\rm seed}'}{0.5}\right)^{\frac{1}{2}}
\\
&\quad\times
\left(\frac{\nu_{\rm obs}}{1.4\,{\rm GHz}}\right)^{\frac{1}{4}}
\frac{\Gamma_{\mathrm{d},0.5}^{\frac{25}{6}}
\Gamma_{\infty,2.5}^{\frac{1}{4}}R_6
\Theta_{-2}^{\frac{1}{2}}\xi_5^{\frac{5}{4}}}
{B_{\mathrm{p},14.3}^{\frac{1}{2}}}.
\end{aligned}
\end{equation}

We next compare the FRB luminosity with the X-ray burst luminosity. Compton drag causes the terminal Lorentz factor of the fireball outflow to depend on the polar angle $\theta$. The flow reaches $\Gamma_\infty$ only within $\theta\lesssim1/\Gamma_\infty$, whereas at larger angles the Lorentz factor is limited to $\Gamma\sim1/\sin\theta$. We further approximate that most of the kinetic energy carried by the relativistic outflow is contained within the $\sim1/\Gamma_\infty$ cone. The true FRB luminosity of this region is then estimated as 
\begin{equation}
\label{eq:frb-luminosity-true}
\begin{aligned}
L_{\rm FRB}^{\rm true}
&\sim\frac{L_{\rm FRB}^{\rm iso}}{4\Gamma_\infty^2}
\\
&=1.2\times10^{36}\,{\rm erg}\,{\rm s}^{-1}\left(\frac{\eta_{\rm seed}'}{0.5}\right)^{\frac{1}{2}}\\
&\quad\times
\left(\frac{\nu_{\rm obs}}{1.4\,{\rm GHz}}\right)^{\frac{1}{4}}
\frac{\Gamma_{\mathrm{d},0.5}^{\frac{25}{6}}
R_6\Theta_{-2}^{\frac{1}{2}}\xi_5^{\frac{5}{4}}}
{B_{\mathrm{p},14.3}^{\frac{1}{2}}
\Gamma_{\infty,2.5}^{\frac{7}{4}}}.
\end{aligned}
\end{equation}
The ratio of the true FRB luminosity to the true X-ray burst luminosity is therefore expressed as
\begin{equation}
\label{eq:frb-xray-efficiency}
\begin{aligned}
\epsilon_{\rm FRB}^{\mathrm{X}}
&\equiv \frac{L_{\rm FRB}^{\rm true}}{L_{\mathrm{X}}}
\sim 1.2\times10^{-5}\left(\frac{\eta_{\rm seed}'}{0.5}\right)^{\frac{1}{2}}\\
&\quad\times
\left(\frac{\nu_{\rm obs}}{1.4\,{\rm GHz}}\right)^{\frac{1}{4}}
\frac{\Gamma_{\mathrm{d},0.5}^{\frac{25}{6}}
R_6\Theta_{-2}^{\frac{1}{2}}\xi_5^{\frac{5}{4}}}
{B_{\mathrm{p},14.3}^{\frac{1}{2}}
\Gamma_{\infty,2.5}^{\frac{7}{4}}L_{\mathrm{X},41}}.
\end{aligned}
\end{equation}

So far, we have fixed the comoving plasma temperature in the scattering region at $\Theta=10^{-2}$. If the plasma receives little additional heating after photon decoupling, however, it can cool substantially before reaching the scattering onset radius. As a conservative estimate for this case, the comoving plasma temperature at the scattering onset radius is given by
\begin{equation}
\label{eq:conservative-temperature}
\begin{aligned}
\Theta_{\rm con}(r_{\rm sc})
&\sim 4.3\times10^{-8}
\left(\frac{\eta_{\mathrm{seed}}'}{0.5}\right)
\left(\frac{\nu_{\mathrm{obs}}}{1.4\,\mathrm{GHz}}\right)^{\frac{1}{2}}
\\
&\quad\times
\frac{T_{\mathrm{C},8}
L_{\mathrm{X},41}^{\frac{5}{12}}
\Gamma_{\mathrm{d},0.5}^{\frac{41}{36}}
\xi_5^{\frac{1}{2}}
}{
R_6^{\frac{5}{12}}
\Gamma_{\infty,2.5}^{\frac{13}{6}}
B_{\mathrm{p},14.3}
},
\end{aligned}
\end{equation}
where $T_{\mathrm{C}}$ is the Compton temperature defined in Eq.~\eqref{eq:Compton_cutoff}.
The detailed derivation is presented in the Appendix \ref{app:conservative-temperature}. In brief, Comptonization regulates the comoving plasma temperature for some time after the photons decouple from the $e^\pm$ plasma. As the outflow subsequently accelerates and expands, the Compton relaxation time eventually exceeds the comoving dynamical time, after which the temperature evolution is dominated by adiabatic cooling. Applying this thermal history to the standard fireball case gives the temperature at $r_{\rm sc}$ in Eq.~\eqref{eq:conservative-temperature}.

Using this conservative temperature in Eq. \eqref{eq:conservative-temperature}, the FRB isotropic luminosity is estimated as
\begin{equation}
\label{eq:frb-luminosity-conservative}
\begin{aligned}
L_{{\rm FRB},{\rm con}}^{\rm iso}
&=1.0\times10^{39}\,{\rm erg}\,{\rm s}^{-1}
\left(\frac{\eta_{\rm seed}'}{0.5}\right)
\left(\frac{\nu_{\rm obs}}{1.4\,{\rm GHz}}\right)^{\frac{1}{2}}
\\
&\quad\times
\frac{T_{\mathrm{C},8}^{\frac{1}{2}}L_{\mathrm{X},41}^{\frac{5}{24}}
\Gamma_{\mathrm{d},0.5}^{\frac{341}{72}}R_6^{\frac{19}{24}}\xi_5^{\frac{3}{2}}}
{\Gamma_{\infty,2.5}^{\frac{5}{6}}B_{\mathrm{p},14.3}}.
\end{aligned}
\end{equation}
This value is substantially smaller than that obtained for a fixed temperature of $\Theta=10^{-2}$ in Eq. \eqref{eq:frb-luminosity-saturated}.

These results show that the conservative pair density corresponding to $\xi=1$ is insufficient to reproduce the luminosity $\sim10^{38}\,{\rm erg}\,{\rm s}^{-1}$ of FRB 20200428 even for the fixed temperature $\Theta=10^{-2}$ from Eq. \eqref{eq:frb-luminosity-saturated}, and adopting the conservative temperature makes the discrepancy more severe. Increasing the pair loading raises the kinetic luminosity and can, in principle, reach the luminosities of bright-end extragalactic FRBs. In the next section, we consider a physical upper limit on the pair loading set by local pair freeze out and evaluate the corresponding FRB luminosity. Possible physical origins of the additional pair loading and plasma heating are discussed in Sec.~\ref{sec:plasma-supply-thermal-evolution}.

\subsection{Local Freeze-out Limit on the Pair Density}
\label{sec:local-freezeout-limit}

Additional pair loading after freeze out can increase the plasma density and thereby enhance both the kinetic luminosity of the fireball outflow and the resulting FRB luminosity. In this section, we adopt the local freeze out density limited by pair annihilation at $r_{\mathrm{sc}}$ as an upper limit on the plasma density. We show that, near this limit, the FRB isotropic luminosity can reach $\sim10^{45}\,{\rm erg}\,{\rm s}^{-1}$, comparable to the luminosities of bright-end extragalactic FRBs.

We consider the case in which the $e^\pm$ plasma freezes out at the scattering radius $r=r_{\mathrm{sc}}$. Possible mechanisms that can supply such a high density pair plasma are discussed in Sec.~\ref{sec:discussion}. Increasing the plasma density enhances the growth of induced Compton scattering because Eq.~\eqref{eq:neutral-efoldings} gives $N_{\mathrm{N}}\propto\xi$. Consequently, Eq.~\eqref{eq:scattering-onset-radius} gives $r_{\mathrm{sc}}\propto\xi^{-\frac{1}{4}}$, so that the scattering radius moves inward as the plasma density increases. The freeze out condition and the scattering onset condition must therefore be solved self-consistently.

The pair annihilation time in the comoving frame is given by Eq.~\eqref{eq:annihilation-time}. Equating this time with the comoving dynamical time in Eq.~\eqref{eq:dynamical-time-decoupling} gives the local pair density at which annihilation freezes out at radius $r$ as
\begin{equation}
\label{eq:local-freeze-density}
n_{\pm,\mathrm{fo}}'(r)
\sim \frac{16}{3}
\frac{\Gamma(r)}{\sigma_{\mathrm{T}}r}.
\end{equation}
The corresponding density enhancement factor relative to the conservative fireball density is obtained from Eqs.~\eqref{eq:standard-pair-density} and \eqref{eq:local-freeze-density} as
\begin{equation}
\label{eq:max-density-enhancement}
\begin{aligned}
\xi_{\max}
= \frac{n_{\pm,\mathrm{fo}}'(r)}
{n_{\pm,{\rm con}}'(r)}= \frac{\Gamma^2r^2}
{\Gamma_{\mathrm{d}}^{\frac{10}{3}}R^2}.
\end{aligned}
\end{equation}

Since the density enhancement also changes the scattering onset radius, $\xi_{\max}$ and $r_{\rm sc}$ must be determined self-consistently. Eliminating $\xi_{\max}$ between Eqs.~\eqref{eq:max-density-enhancement} and \eqref{eq:scattering-onset-radius}, we obtain the scattering onset radius in the local freeze out case as
\begin{equation}
\label{eq:scattering-radius-max-density}
\begin{aligned}
r_{\rm sc}(\xi_{\max})
&=1.4\times10^8\,{\rm cm}
\left(\frac{\eta_{\rm seed}'}{0.5}\right)^{-\frac{1}{3}}
\left(\frac{\nu_{\rm obs}}{1.4\,{\rm GHz}}\right)^{-\frac{1}{6}}
\\
&\quad\times
\Gamma_{\infty,2.5}^{\frac{1}{6}}
B_{\mathrm{p},14.3}^{\frac{1}{3}}R_6\sim r_{\mathrm{acc}}.
\end{aligned}
\end{equation}
For the fiducial parameters, the scattering onset radius is comparable to the coasting radius given by Eq.~\eqref{eq:acceleration-radius}. For simplicity, we approximate the Lorentz factor at the scattering radius as $\Gamma(r_{\mathrm{sc}})\simeq\Gamma_\infty$.\footnote{For the fiducial parameters, this approximation overestimates the FRB luminosity given by Eq.~\eqref{eq:frb-luminosity-freeze-iso} by a factor of a few. This uncertainty is small compared with the range of more than ten orders of magnitude in FRB luminosity shown in Fig.~\ref{fig:fireball_outflow_luminosity}. A more accurate estimate would require evaluating the FRB luminosity using the local Lorentz factor at the scattering onset radius when scattering begins during the acceleration phase.}

The density enhancement factor at this scattering onset radius is obtained by substituting Eq.~\eqref{eq:scattering-radius-max-density} into Eq.~\eqref{eq:max-density-enhancement}, which gives
\begin{equation}
\label{eq:max-density-enhancement-evaluated}
\begin{aligned}
\xi_{\max}(r_{\rm sc})
&=4.7\times10^7
\left(\frac{\eta_{\rm seed}'}{0.5}\right)^{-\frac{2}{3}}
\left(\frac{\nu_{\rm obs}}{1.4\,{\rm GHz}}\right)^{-\frac{1}{3}}
\\
&\quad\times
\frac{\Gamma_{\infty,2.5}^{\frac{7}{3}}
B_{\mathrm{p},14.3}^{\frac{2}{3}}}
{\Gamma_{\mathrm{d},0.5}^{\frac{10}{3}}}.
\end{aligned}
\end{equation}
The corresponding density of the $e^\pm$ plasma is
\begin{equation}
\label{eq:local-freeze-density-scattering}
\begin{aligned}
n_{\pm,\mathrm{fo}}'(r_{\rm sc})
&=1.8\times10^{19}\,{\rm cm}^{-3}
\left(\frac{\eta_{\rm seed}'}{0.5}\right)^{\frac{1}{3}}
\left(\frac{\nu_{\rm obs}}{1.4\,{\rm GHz}}\right)^{\frac{1}{6}}
\\
&\quad\times
\frac{\Gamma_{\infty,2.5}^{\frac{5}{6}}}{R_6B_{\mathrm{p},14.3}^{\frac{1}{3}}}.
\end{aligned}
\end{equation}

We next estimate the FRB luminosity associated with this local freeze out density. Substituting Eqs.~\eqref{eq:max-density-enhancement-evaluated} and \eqref{eq:scattering-radius-max-density} into Eq.~\eqref{eq:kinetic-luminosity-scattering}, the isotropic kinetic luminosity at the scattering onset radius is given by
\begin{equation}
\label{eq:kinetic-luminosity-freeze-iso}
\begin{aligned}
L_{{\rm kin},\mathrm{fo}}^{\rm iso}(r_{\rm sc})
&=1.1\times10^{46}\,{\rm erg}\,{\rm s}^{-1}
\left(\frac{\eta_{\rm seed}'}{0.5}\right)^{-\frac{1}{3}}
\\
&\quad\times
\left(\frac{\nu_{\rm obs}}{1.4\,{\rm GHz}}\right)^{-\frac{1}{6}}
R_6\Gamma_{\infty,2.5}^{\frac{19}{6}}B_{\mathrm{p},14.3}^{\frac{1}{3}}.
\end{aligned}
\end{equation}
For a fixed comoving plasma temperature of $\Theta=10^{-2}$, the FRB isotropic luminosity is estimated as
\begin{equation}
\label{eq:frb-luminosity-freeze-iso}
\begin{aligned}
L_{{\rm FRB},\mathrm{fo}}^{\rm iso}
&=\Theta^{\frac{1}{2}}L_{{\rm kin},\mathrm{fo}}^{\rm iso}
\\
&=1.1\times10^{45}\,{\rm erg}\,{\rm s}^{-1}
\left(\frac{\eta_{\rm seed}'}{0.5}\right)^{-\frac{1}{3}}
\left(\frac{\nu_{\rm obs}}{1.4\,{\rm GHz}}\right)^{-\frac{1}{6}}
\\
&\quad\times
R_6\Gamma_{\infty,2.5}^{\frac{19}{6}}
B_{\mathrm{p},14.3}^{\frac{1}{3}}\Theta_{-2}^{\frac{1}{2}}.
\end{aligned}
\end{equation}
This luminosity is comparable to the bright end of the observed extragalactic FRB population. The second CHIME/FRB catalog gives a characteristic upper energy scale of $E_{\mathrm{max}}^{\mathrm{iso}}\sim1.2\times10^{42}\,{\rm erg}$ \citep{Shah_2026}. For a representative burst duration of $\Delta t\sim10^{-3}\,\mathrm{s}$, this energy corresponds to $L_{\mathrm{max}}^{\mathrm{iso}}\sim10^{45}\,{\rm erg}\,{\rm s}^{-1}$, comparable to the value in Eq.~\eqref{eq:frb-luminosity-freeze-iso}. Thus, if the pair density approaches the local freeze out limit, induced Compton upscattering can produce the bright-end extragalactic FRBs.

\subsection{Mapping X-ray Burst Luminosity to FRB}
\label{sec:xray-burst-to-frb-emission}

\begin{figure*}
\centering
\includegraphics[width=\textwidth]{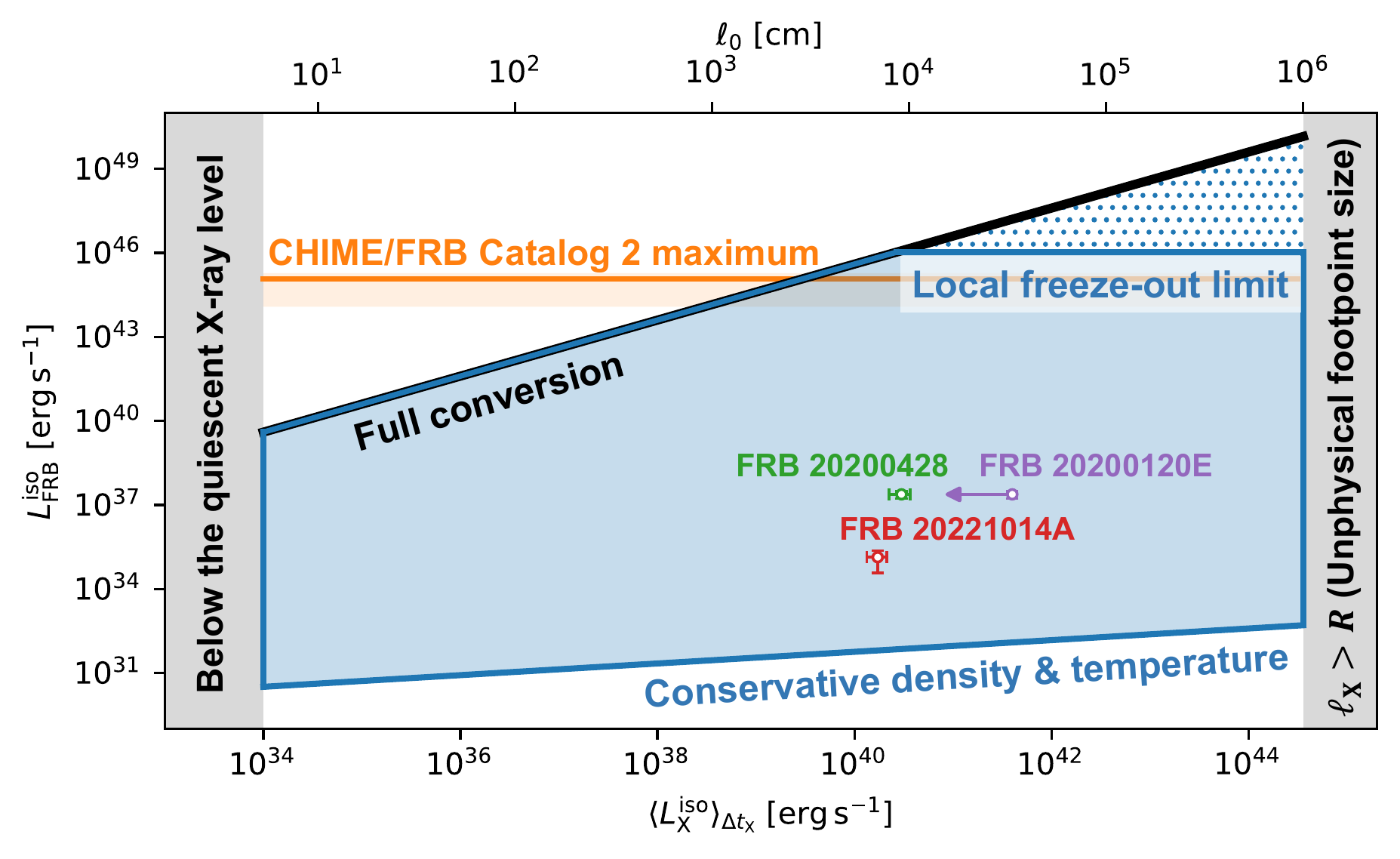}
\caption{\justifying
The lower horizontal axis shows the X-ray burst isotropic luminosity, and the vertical axis shows the FRB isotropic luminosity. The upper horizontal axis shows the footpoint radius $\ell_0$ of the fireball at the magnetar surface corresponding to each X-ray luminosity, as determined from Eq.~\eqref{eq:initial-footpoint-radius}. The blue region represents the range of X-ray and FRB luminosities obtained by varying the plasma density and temperature considered in this work. The lower boundary corresponds to the conservative fireball case. The upper boundary corresponds to the local freeze out limit for the pair density, formally extrapolated to $\Theta=1$. The blue dotted region above this boundary indicates a formal region that would remain energetically allowed if plasma densities exceeding the local freeze out limit could be realized. The black solid line shows the upper limit set by the energy budget when the entire X-ray luminosity is converted into the true kinetic luminosity of the outflow and subsequently into the true FRB luminosity. The red, green, and purple points show observed FRBs with associated X-ray bursts or simultaneous X-ray upper limits. For FRB~20200120E, shown in purple, we use the X-ray upper limit obtained from simultaneous observations. The orange band represents the luminosity obtained by converting the bright end energy scale of one-off FRBs inferred from the second CHIME/FRB catalog using a representative burst duration. The gray region on the left corresponds to X-ray luminosities below the typical persistent X-ray luminosity of magnetars and is outside the range considered in this work. The gray region on the right requires $\ell_0>R$ for a fixed temperature of $T_0\sim10^9\,\mathrm{K}$ and is therefore incompatible with the fireball geometry assumed at the magnetar surface.
}
\label{fig:fireball_outflow_luminosity}
\end{figure*}

Fig.~\ref{fig:fireball_outflow_luminosity} summarizes one of the main results of this work by showing the range of FRB isotropic luminosities that can be produced for a given X-ray burst luminosity. According to Eq.~\eqref{eq:initial-footpoint-radius}, the X-ray luminosity can be varied in two ways. One is to vary the footpoint radius $\ell_0$ while keeping the blackbody temperature $T_0$ fixed. The other is to vary $T_0$ while keeping $\ell_0$ fixed. We adopt the former approach and investigate the dependence on the X-ray luminosity by varying $\ell_0$ at fixed $T_0$.\footnote{We vary $\ell_0$ rather than $T_0$ to minimize changes in the fireball dynamics. As shown in Eq.~\eqref{eq:pair-equilibrium-density}, varying $T_0$ changes the equilibrium pair density of the $e^\pm$ plasma exponentially. Moreover, if the temperature becomes relativistic, the fireball dynamics must be reconsidered. A systematic investigation of this temperature dependence is left for future work. In contrast, varying only $\ell_0$ at fixed $T_0$ has a relatively weak effect on the fireball dynamics. Specifically, the decoupling radius and the Lorentz factor at that radius, given by Eq.~\eqref{eq:decoupling-radius-lorentz-factor}, vary by less than a factor of two over the range considered in this work. We neglect these variations below. This approximation can change the FRB luminosity by a factor of a few but does not change its order of magnitude.}

The upper and lower boundaries of the blue-colored region represent a formal upper estimate and a conservative lower estimate, respectively, of the FRB luminosity from the polar fireball outflow. The lower boundary, labeled ``conservative density \& temperature,'' is obtained by setting $\xi=1$ in Eq.~\eqref{eq:frb-luminosity-conservative}. The upper boundary, labeled ``local freeze-out limit,'' is obtained from Eq.~\eqref{eq:frb-luminosity-freeze-iso} by formally setting $\Theta=1$. Thus, increasing the X-ray luminosity alone does not allow the FRB luminosity to grow arbitrarily. The maximum FRB luminosity is constrained by the pair density, the plasma temperature, and the available energy budget.

We next derive the remaining boundaries shown in Fig.~\ref{fig:fireball_outflow_luminosity}. The black solid line labeled ``full conversion'' represents an
extreme upper bound on the global energy budget. Here we assume
that the total X-ray burst power can be supplied to the narrow
relativistic outflow within $\theta\lesssim1/\Gamma_\infty$ and
subsequently converted into kinetic and FRB luminosity. Under this
maximal energy transfer assumption,
\begin{equation}
\label{eq:luminosity-equality-boundary}
L_{\mathrm{X}} = L_{\rm kin}^{\mathrm{true}}=L_{\rm FRB}^{\mathrm{true}}.
\end{equation}
Therefore, an FRB luminosity above this boundary cannot be produced if the X-ray luminosity is the available energy source. Converting Eq.~\eqref{eq:luminosity-equality-boundary} into the FRB isotropic luminosity gives
\begin{equation}
\label{eq:frb-isotropic-upper-boundary}
\begin{aligned}
L_{\rm FRB}^{\rm iso}
&=4\Gamma_\infty^2L_{\mathrm{X}}
\\
&=4\times10^{46}\,{\rm erg}\,{\rm s}^{-1}
\Gamma_{\infty,2.5}^2L_{\mathrm{X},41}.
\end{aligned}
\end{equation}
which provides an upper limit imposed by energy conservation.

The assumed fireball geometry also places an upper limit on the X-ray luminosity itself. According to Eq.~\eqref{eq:initial-footpoint-radius}, the X-ray luminosity increases with the footpoint radius $\ell_0$ when $T_0$ is fixed. We adopt $\ell_0=R$ as the maximum footpoint size. The corresponding maximum X-ray luminosity is given by
\begin{equation}
\label{eq:max-xray-luminosity}
L_{\mathrm{X},\max}
=3.6\times10^{44}\,{\rm erg}\,{\rm s}^{-1}
T_{0,9}^4R_6^2.
\end{equation}
We adopt Eq.~\eqref{eq:max-xray-luminosity} as the upper limit on the X-ray luminosity considered in this work.

\begin{table*}[t]
\caption{
Observed radio and X-ray quantities adopted for the three events shown in
Fig.~\ref{fig:fireball_outflow_luminosity}.
}
\label{tab:observed_frbs}
\centering
\scriptsize
\setlength{\tabcolsep}{3.5pt}
\renewcommand{\arraystretch}{1.25}
\begin{tabular}{lcccccccc}
\hline\hline
Event
&
Distance
&
Radio band
&
$\Delta t_{\mathrm{FRB}}$
&
$L_{\mathrm{FRB}}^{\mathrm{iso}}$\textsuperscript{a}
&
X-ray band
&
$\Delta t_X$
&
$\langle L_X^{\mathrm{iso}}\rangle_{\Delta t_X}$\textsuperscript{c}
&
References
\\
&
&
$[\mathrm{MHz}]$
&
$[\mathrm{ms}]$
&
$[\mathrm{erg\,s^{-1}}]$
&
$[\mathrm{keV}]$
&
$[\mathrm{s}]$
&
$[\mathrm{erg\,s^{-1}}]$
&
\\
\hline
FRB~20200428
&
$6.6\pm0.7~\mathrm{kpc}$
&
$1281$--$1468$
&
$0.61$
&
$(2.4\pm0.7)\times10^{37}$
&
$20$--$200$
&
$0.010$
&
$3.0^{+0.69}_{-0.76}\times10^{40}$
&
(1)--(3)
\\

FRB~20221014A
&
$6.6\pm0.7~\mathrm{kpc}$
&
$400$--$800$
&
$1.48$
&
$(1.37\pm0.99)\times10^{35}$
&
$1$--$250$
&
$0.23$
&
$1.72^{+0.41}_{-0.39}\times10^{40}$\textsuperscript{d}
&
(3)--(5)
\\

FRB~20200120E~(B4)
&
$3.63\pm0.34~\mathrm{Mpc}$
&
$1255$--$1505$
&
$0.117$
&
$(2.4\pm0.7)\times10^{37}$\textsuperscript{b}
&
$0.5$--$10$
&
$0.1$
&
$<4.0\times10^{41}$
&
(6)
\\
\hline\hline
\end{tabular}

\vspace{3pt}
\begin{minipage}{0.99\textwidth}
\raggedright
\scriptsize

\textsuperscript{a}
We estimate the isotropic energy and luminosity as
$E_{\mathrm{FRB}}^{\mathrm{iso}}
=4\pi d_\mathrm{L}^2\mathcal{F}_{\nu}\Delta\nu_{\mathrm{obs}}/(1+z)$
and
$L_{\mathrm{FRB}}^{\mathrm{iso}}
=4\pi d_L^2\mathcal{F}_{\nu}\Delta\nu_{\mathrm{obs}}/
\Delta t_{\mathrm{FRB,obs}}$,
respectively, where $\mathcal{F}_{\nu}$ is the spectral fluence density.
For FRB~20200428 and FRB~20221014A, the radio emission reaches the edge of
the observing band, and we therefore conservatively adopt the full instrumental
bandwidth as $\Delta\nu_{\mathrm{obs}}$.
The resulting radio luminosities may underestimate the intrinsic values if the
emission extends beyond the observed frequency range.

\textsuperscript{b}
FRB~20200120E~(B4) is contained within the observing band.
We nevertheless follow the published value of
$E_{\mathrm{FRB}}^{\mathrm{iso}}$ and estimate the luminosity as
$L_{\mathrm{FRB}}^{\mathrm{iso}}
=E_{\mathrm{FRB}}^{\mathrm{iso}}/\Delta t_{\mathrm{FRB}}$.

\textsuperscript{c}
The X-ray luminosity is averaged over the adopted interval $\Delta t_{\mathrm{X}}$,
$\langle L_{\mathrm{X}}^{\mathrm{iso}}\rangle_{\Delta t_{\mathrm{X}}}
\equiv E_{\mathrm{X}}^{\mathrm{iso}}/\Delta t_{\mathrm{X}}$.
Since the adopted X-ray energy bands differ among the three events,
the listed X-ray luminosities may underestimate the intrinsic luminosities
by different amounts.

\textsuperscript{d}
The isotropic X-ray energy of FRB~20221014A was originally
estimated assuming a distance of $9~\mathrm{kpc}$.
For consistency with FRB~20200428, we rescale the energy to
$d_L=6.6~\mathrm{kpc}$ using
$E_{\mathrm{X}}^{\mathrm{iso}}\propto d_L^2$
and calculate the corresponding X-ray luminosity.
Although the radio burst is temporally aligned with the P2 X-ray peak,
the X-ray flux was not analyzed separately for the individual peaks.
We therefore estimate the luminosity using the duration and energy of the
entire X-ray burst.

References:
(1)~\citet{Bochenek2020-ev};
(2)~\citet{Mereghetti2020-ka};
(3)~\citet{Zhou2020-fb};
(4)~\citet{Wang2026-bm};
(5)~\citet{2023arXiv231016932G};
(6)~\citet{Pearlman2025-pa}.
\end{minipage}
\end{table*}

We compare the predicted luminosity range with three representative observational constraints in Fig.~\ref{fig:fireball_outflow_luminosity}. FRB~20200428 \citep{Bochenek2020-ev} and FRB~20221014A \citep{Wang2026-bm} are Galactic events for which a strong association between the radio bursts and X-ray bursts has been established. FRB~20200120E \citep{Pearlman2025-pa} is included as a representative extragalactic event with a stringent upper limit from simultaneous X-ray observations. FRB~20200428 was detected by both CHIME/FRB \citep{Andersen2020-iz} and STARE2 \citep{Bochenek2020-ev}. We use the STARE2 band because it contained the larger radio energy. The adopted observational quantities are summarized in Tab.~\ref{tab:observed_frbs}.

The orange band in Fig.~\ref{fig:fireball_outflow_luminosity} represents the luminosity corresponding to the characteristic cutoff energy at the bright end of the energy distribution of one-off FRBs inferred from the second CHIME/FRB catalog, $E_{\mathrm{max}}^{\mathrm{iso}}\sim1.2\times10^{42}\,\mathrm{erg}$ \citep{Shah_2026}. We convert this energy scale into a luminosity scale by assuming a representative radio duration of $\Delta t_{\mathrm{R}}=1\,\mathrm{ms}$.

Fig.~\ref{fig:fireball_outflow_luminosity} shows that the present scenario can span a broad luminosity range from Galactic FRBs to bright extragalactic FRBs. The observed luminosities of FRB~20200428 and FRB~20221014A lie within the range represented by the blue region. The simultaneous X-ray upper limit for FRB~20200120E also does not exclude the present scenario. Moreover, the luminosity scale inferred from the bright end of the one-off FRB energy distribution in the second CHIME/FRB catalog is comparable to the FRB luminosity obtained when the pair density approaches the local freeze out limit. Thus, at least in terms of luminosity, both relatively faint Galactic FRBs and bright extragalactic FRBs can be accommodated within the same polar fireball outflow framework.

\section{Small pair loading region and spectral bandwidth}
\label{sec:small}

\begin{figure*}
  \centering
  \includegraphics[width=\textwidth]{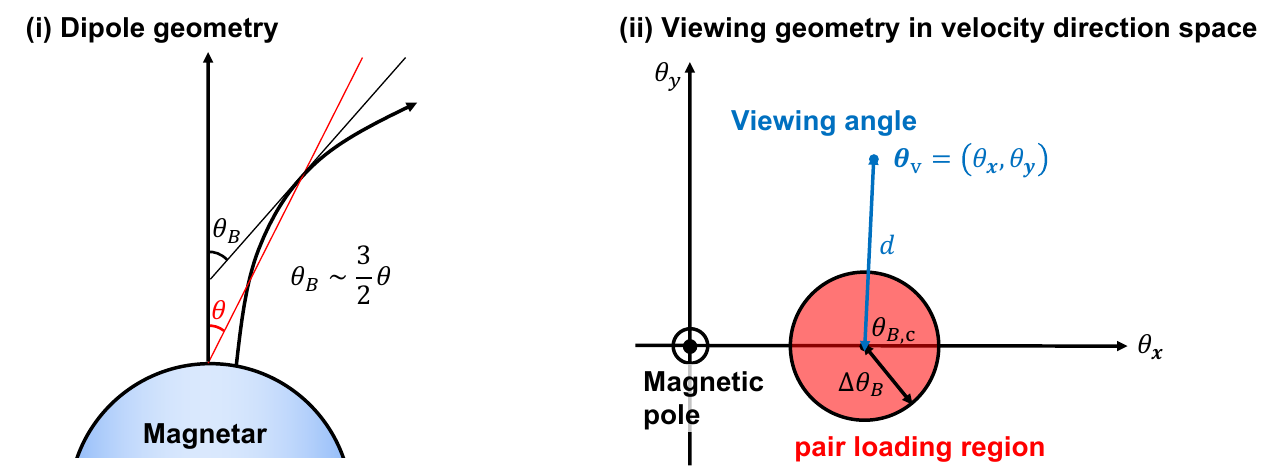}
  \caption{\justifying Definition of the angular parameters used for the small pair loading region.
(i) We denote by $\theta$ the polar angle of the plasma position measured from the magnetic axis and by $\theta_B$ the angle of the plasma velocity direction, which follows the local dipole magnetic field. In general, $\theta_B\neq\theta$.
(ii) The red region shows the angular extent of the velocity directions associated with the pair loading region. The blue point $\bm{\theta}_{\mathrm{v}}$ denotes the two-dimensional viewing-angle vector measured from the magnetic axis. The dimensionless angular separation between $\bm{\theta}_{\mathrm{v}}$ and the center of the velocity direction region $\theta_{B,\mathrm{c}}$ is denoted by $d$ and defined in Eq.~\eqref{eq:viewing-offset-parameter}. A viewing offset of $d\lesssim 1$ is typical for an observer within the relativistic beaming cone.
  }
  \label{fig:small_region}
\end{figure*}

\begin{figure*}
  \centering
  \includegraphics[width=\textwidth]{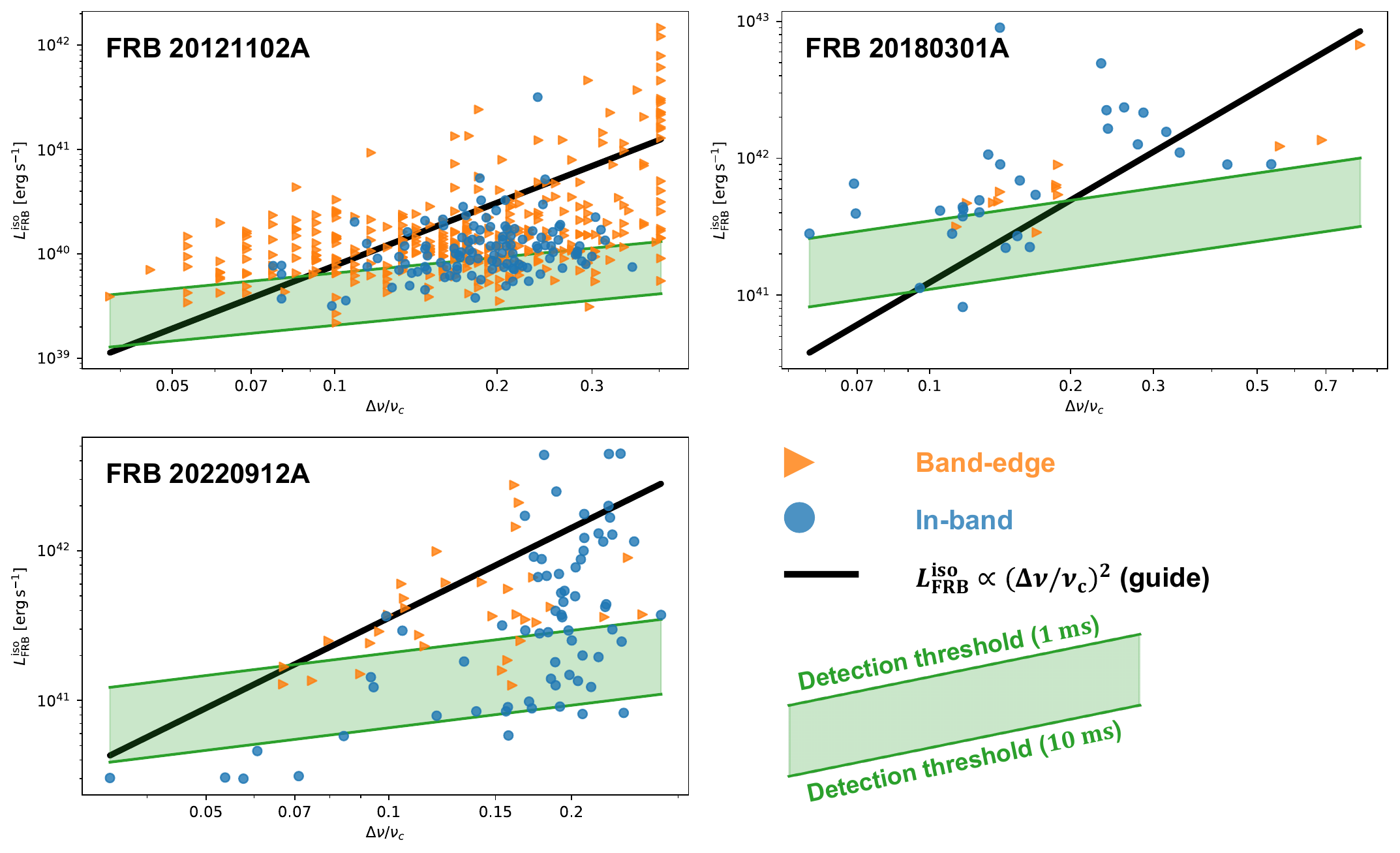}
  \caption{\justifying Distributions of the fractional spectral bandwidth and FRB isotropic luminosity for bursts from the three repeating FRBs.
Here, $\nu_{\mathrm{c}}$ and $\Delta\nu$ denote the central frequency and
spectral bandwidth of each burst, respectively.
Blue circles represent bursts whose spectral extents are judged to be
contained within the observing band, while orange triangles represent
bursts whose spectral extents are truncated or uncertain near a band edge.
The bandwidths of the orange triangle sample may therefore underestimate the
intrinsic values.
The black solid lines show
$L_{\rm FRB}^{\rm iso}\propto(\Delta\nu/\nu_{\mathrm{c}})^2$
as a guide.
The green bands show the detection thresholds based on the instrumental sensitivities reported in the
corresponding references.
The upper and lower boundaries correspond to burst durations of
$1\,{\rm ms}$ and $10\,{\rm ms}$, respectively.
For FRB 20121102A \citep{2022MNRAS.515.3577H} and
FRB 20180301A \citep{2023MNRAS.526.3652K},
we calculate $\nu_{\mathrm{c}}$ and $\Delta\nu$ from the lower and upper
spectral edges tabulated in the respective references as
$\nu_{\mathrm{c}}=(\nu_{\mathrm{high}}+\nu_{\mathrm{low}})/2$ and
$\Delta\nu=\nu_{\mathrm{high}}-\nu_{\mathrm{low}}$.
For FRB 20220912A \citep{2024MNRAS.52710425S},
we directly use the tabulated values of $\nu_{\mathrm{c}}$ and $\Delta\nu$.
Since individual band edge classifications are not provided in the
tabulated data for FRB 20220912A, we classify bursts as a band edge
case when either
$\nu_{\mathrm{c}}-\Delta\nu/2$ or
$\nu_{\mathrm{c}}+\Delta\nu/2$
lies outside the corresponding observing band.
  }
  \label{fig:fireball_outflow_bandwidth}
\end{figure*}


In this section, we show that the finite angular size of a localized pair loading region can produce a correlation between the FRB luminosity and the fractional spectral bandwidth. For simplicity, we assume a monochromatic intrinsic frequency $\nu'_1$ and a single bulk Lorentz factor $\Gamma$. The bandwidth derived below therefore represents the minimum width arising from the geometrical effect. We consider plasma flowing along dipolar magnetic field lines near the magnetic pole and assume that $e^\pm$ plasma is locally supplied in excess within a circular region of angular radius $\theta_{\mathrm{p}}$ centered at the polar angle $\theta_{\mathrm{c}}$. As illustrated in Fig.~\ref{fig:small_region}(i), the plasma velocity follows the local dipolar magnetic field and is directed at an angle $\theta_B$ from the magnetic axis. The radial and polar components of the dipole magnetic field are
\begin{equation}
\label{eq:dipole-field-components}
\begin{aligned}
B_r
&=
B_{\mathrm{p}}
\frac{R^3}{r^3}
\cos\theta,
\\
B_\theta
&=
B_{\mathrm{p}}
\frac{R^3}{2r^3}
\sin\theta,
\end{aligned}
\end{equation}
so that
$\tan(\theta_B-\theta)=B_\theta/B_r=(1/2)\tan\theta$.
Near the magnetic pole, where $\theta_B-\theta\ll1$ and $\theta\ll1$, this relation reduces to
\begin{equation}
\label{eq:dipole-field-angle}
\theta_B
\simeq
\frac{3}{2}\theta.
\end{equation}
A circular pair loading region in physical space therefore corresponds to the red region in velocity direction space shown in Fig.~\ref{fig:small_region}(ii). Its center and angular radius in velocity direction space are
\begin{equation}
\label{eq:pair-loading-velocity-center}
\bm{\theta}_{B,\mathrm{c}}
\equiv
\frac{3}{2}\bm{\theta}_{\mathrm{c}},
\end{equation}
and
\begin{equation}
\label{eq:pair-loading-velocity-radius}
\Delta\theta_B
\equiv
\frac{3}{2}\theta_{\mathrm{p}},
\end{equation}
respectively.

We denote the viewing direction measured from the magnetic axis by the two-dimensional angular vector $\bm{\theta}_{\mathrm{v}}$. We consider a pair loading region much smaller than the relativistic beaming cone,
\begin{equation}
\label{eq:small_condition}
\Gamma\Delta\theta_B\ll1.
\end{equation}
We define the dimensionless angular separation between the viewing direction and the center of the velocity region as
\begin{equation}
\label{eq:viewing-offset-parameter}
d
\equiv
\Gamma
\left|
\bm{\theta}_{\mathrm{v}}
-
\bm{\theta}_{B,\mathrm{c}}
\right|.
\end{equation}
We focus on viewing directions satisfying
\begin{equation}
\label{eq:geometry_condition}
\Gamma\Delta\theta_B\ll d\lesssim1.
\end{equation}
The lower bound places the line of sight outside the small velocity region, while the upper bound keeps it within the relativistic beaming cone, beyond which the observed emission is strongly suppressed. Since $\Delta\theta_B\ll1/\Gamma$, the probability that the line of sight lies directly within the velocity region is correspondingly small. In this geometry, a smaller $\Delta\theta_B$ produces a narrower observed bandwidth, leading to the luminosity--bandwidth relation derived below.

We next estimate the observed frequency range produced by the spread of velocity directions within the pair loading region. Let $\psi$ be the angle between the direction of the local bulk velocity and the line of sight to the observer. The observed frequency is given by the Doppler factor as
\begin{equation}
\label{eq:observed-frequency-viewing-angle}
\begin{aligned}
\nu_{\mathrm{obs}}
&=
\frac{\nu_1'}
{\Gamma\left(1-\beta\cos\psi\right)}
\\
&\simeq
\frac{2\Gamma}
{1+\Gamma^2\psi^2}
\nu_1',
\qquad
\Gamma\gg1,
\quad
\psi\ll1,
\end{aligned}
\end{equation}
and we define the fractional spectral bandwidth as
\begin{equation}
\label{eq:fractional-spectral-bandwidth}
\frac{\Delta\nu}{\nu_{\mathrm{c}}}
\equiv
\frac{
2\left(\nu_{\max}-\nu_{\min}\right)
}{
\nu_{\max}+\nu_{\min}
},
\end{equation}
where $\nu_{\max}$ and $\nu_{\min}$ are the maximum and minimum observed frequencies produced within the pair loading region.

As illustrated in Fig.~\ref{fig:small_region}(ii), the minimum angular separation between the viewing direction and the velocity region is
\begin{equation}
\label{eq:min-viewing-angle}
\Gamma\psi_{\min}
=
d-\Gamma\Delta\theta_B.
\end{equation}
The corresponding maximum separation is
\begin{equation}
\label{eq:max-viewing-angle}
\Gamma\psi_{\max}
=
d+\Gamma\Delta\theta_B.
\end{equation}
Using Eq.~\eqref{eq:observed-frequency-viewing-angle}, the maximum and minimum observed frequencies are therefore expressed as
\begin{equation}
\label{eq:max-min-observed-frequency}
\begin{aligned}
\nu_{\max}
&=
\frac{2\Gamma\nu_1'}
{1+\Gamma^2\psi_{\min}^2}
=
\frac{2\Gamma\nu_1'}
{1+\left(d-\Gamma\Delta\theta_B\right)^2},
\\
\nu_{\min}
&=
\frac{2\Gamma\nu_1'}
{1+\Gamma^2\psi_{\max}^2}
=
\frac{2\Gamma\nu_1'}
{1+\left(d+\Gamma\Delta\theta_B\right)^2}.
\end{aligned}
\end{equation}

Substituting Eq.~\eqref{eq:max-min-observed-frequency} into Eq.~\eqref{eq:fractional-spectral-bandwidth} gives
\begin{equation}
\label{eq:bandwidth-pair-loading-size}
\begin{aligned}
\frac{\Delta\nu}{\nu_{\mathrm{c}}}
&=
\frac{
4d\Gamma\Delta\theta_B
}{
1+d^2+\Gamma^2\Delta\theta_B^2
}
\\
&\simeq
\frac{4d}{1+d^2}\Gamma\Delta\theta_B
\propto \Gamma\Delta\theta_B,
\qquad
\Gamma\Delta\theta_B\ll1.
\end{aligned}
\end{equation}
Thus, in the limit of a small pair loading region, the fractional spectral bandwidth is proportional to its angular size in velocity direction space. Combining Eqs.~\eqref{eq:pair-loading-velocity-radius} and \eqref{eq:bandwidth-pair-loading-size}, the physical radius of the pair loading region is expressed as
\begin{equation}
\label{eq:pair-loading-radius-bandwidth}
r\theta_{\mathrm{p}}
\simeq
\frac{r}{\Gamma}
\frac{1+d^2}{6d}
\frac{\Delta\nu}{\nu_{\mathrm{c}}}.
\end{equation}

This relation directly determines the dependence of the FRB luminosity on the fractional spectral bandwidth. If the radiative energy flux per unit area within the pair loading region is given by Eq.~\eqref{eq:saturation-flux}, the true luminosity emitted from a circular region of radius $r\theta_{\mathrm{p}}$ is estimated as
\begin{equation}
\label{eq:true-frb-luminosity-small-region}
\begin{aligned}
L_{\mathrm{FRB}}^{\mathrm{true}}
&=
\pi
\left(r\theta_{\mathrm{p}}\right)^2
F_{\mathrm{sat}}
\\
&\simeq
\pi r^2
\left(
\frac{1+d^2}{6d}
\right)^2
\left(
\frac{\Delta\nu}{\nu_{\mathrm{c}}}
\right)^2
n_\pm'
m_{\mathrm{e}}c^3
\Theta^{\frac{1}{2}}.
\end{aligned}
\end{equation}

Finally, under the condition in Eq.~\eqref{eq:small_condition}, the radiation from each fluid element is concentrated within an angle of order $1/\Gamma$. Neglecting the order unity dependence on the viewing geometry associated with $d\lesssim 1$, the isotropic luminosity is estimated as
\begin{equation}
\label{eq:isotropic-frb-luminosity-bandwidth}
\begin{aligned}
L_{\mathrm{FRB}}^{\mathrm{iso}}
&\simeq
4\Gamma^2
L_{\mathrm{FRB}}^{\mathrm{true}}
\\
&\simeq
4\pi r^2
\left(
\frac{1+d^2}{6d}
\right)^2
\left(
\frac{\Delta\nu}{\nu_{\mathrm{c}}}
\right)^2
\Gamma^2
n_\pm'
m_{\mathrm{e}}c^3
\Theta^{\frac{1}{2}}
\\
&\propto
\left(
\frac{\Delta\nu}{\nu_{\mathrm{c}}}
\right)^2.
\end{aligned}
\end{equation}
Thus, for a sufficiently small pair loading region viewed from a direction outside the velocity region but within the relativistic beaming cone in Eq. \eqref{eq:geometry_condition}, the model predicts a luminosity--bandwidth relation
$L_{\mathrm{FRB}}^{\mathrm{iso}}\propto(\Delta\nu/\nu_{\mathrm{c}})^2$.

The observed burst samples of three repeating FRBs show an apparent tendency for bursts with smaller fractional spectral bandwidths to have lower isotropic luminosities, qualitatively consistent with the prediction above. Fig.~\ref{fig:fireball_outflow_bandwidth} shows $L_{\mathrm{FRB}}^{\mathrm{iso}}$ as a function of $\Delta\nu/\nu_{\mathrm{c}}$ for bursts from FRB 20121102A \citep{2022MNRAS.515.3577H}, FRB 20180301A \citep{2023MNRAS.526.3652K}, and FRB 20220912A \citep{2024MNRAS.52710425S}. The black solid lines show the scaling $L_{\mathrm{FRB}}^{\mathrm{iso}}\propto(\Delta\nu/\nu_{\mathrm{c}})^2$ as a guide. Details of the data analysis and the treatment of the spectral bandwidths are described in the caption of Fig.~\ref{fig:fireball_outflow_bandwidth}.

We emphasize two caveats in interpreting the comparison with the observed bandwidth. First, $\Delta\nu_{\mathrm{obs}}$ enters linearly into the definitions of $E_{\mathrm{FRB}}^{\mathrm{iso}}=4\pi d_{\mathrm{L}}^2\mathcal{F}_{\nu}\Delta\nu_{\mathrm{obs}}/(1+z)$. Therefore, part of any positive correlation with $\Delta\nu/\nu_{\mathrm{c}}$ is built into the observed quantities \citep{2022MNRAS.515.3577H}, and a dependence steeper than linear would provide a more meaningful indication of an additional physical trend. Second, our prediction assumes a monochromatic intrinsic frequency $\nu'_1$ and a single bulk Lorentz factor $\Gamma$. In reality, the low frequency seed waves may be broadband \citep{2020ApJ...897..173B,2026ApJ...998..190Q}. The finite bandwidth of the seed wave can therefore contribute to the bandwidth of the scattered radiation. Variations in $\Gamma$ can introduce additional broadening. The bandwidth estimated here should therefore be regarded as a minimum width. These effects may weaken or wash out the simple relation $L_{\mathrm{FRB}}^{\mathrm{iso}}\propto(\Delta\nu/\nu_{\mathrm{c}})^2$. Accordingly, the black solid line in Fig.~4 should be viewed as an idealized baseline rather than a generic prediction of the model.

\section{Discussion}
\label{sec:discussion}
\subsection{Plasma Supply and Thermal Evolution}
\label{sec:plasma-supply-thermal-evolution}
\subsubsection{Additional Plasma Supply and Outflow Energetics}
Producing bright FRBs in this model requires a plasma density higher than the conservative fireball density in Eq.~\eqref{eq:standard-pair-density}. We parameterize this additional pair loading through the density enhancement factor in Eq.~\eqref{eq:density-enhancement}, without specifying its physical origin. Possible sources include magnetic reconnection \citep{Beloborodov2021-sf}, dissipation of EM waves, and internal shocks produced by collisions between fireball outflows. These processes may generate high energy photons that create additional $e^\pm$ plasma. However, it remains to be determined how efficiently the newly created pairs are incorporated into the polar fireball outflow.

Radiative magnetic reconnection can generate high energy photons and drive efficient $\gamma\gamma$ pair creation \citep{Beloborodov2021-sf}, while $\gamma B$ pair creation may also operate closer to the magnetar surface, where the background magnetic field is stronger. Whether these processes can supply a sufficient number of pairs to the polar outflow and maintain the required density up to the scattering radius remains to be investigated. A self-consistent treatment of pair creation, annihilation, and transport is therefore required to determine the physically attainable value of $\xi$.

Observations of the radio afterglow following the 2004 giant flare from SGR~1806--20 provide an reference for the energetics of magnetar outflows \citep{2005Natur.434.1104G,2005ApJ...634L..89G,2006ApJ...638..391G}. Based on afterglow modeling, \citet{2006ApJ...638..391G} inferred a lower limit of $\sim5\times10^{45}\ {\rm erg}$ on the isotropic kinetic energy of the ejecta \citep{2005ApJ...635..516N}. By comparison, for our fiducial pair loading factor $\xi\sim10^5$, Eq.~\eqref{eq:kinetic-luminosity-scattering} gives $L_{\rm kin}^{\rm iso}\sim10^{42}\ {\rm erg\,s^{-1}}$. For an outflow duration of $\sim0.1\ {\rm s}$, this corresponds to $E_{\rm kin}^{\rm iso}\sim10^{41}\ {\rm erg}$, well below the kinetic energy inferred for the giant flare ejecta. Thus, the overall energy requirement of the fiducial outflow is not exceptionally large compared with energetic magnetar ejecta. However, the giant flare ejecta were inferred to be mildly relativistic and baryon rich. This comparison therefore does not establish whether a pure $e^\pm$ outflow can carry the required kinetic energy while reaching the ultrarelativistic Lorentz factor assumed in our model.

Baryon loading provides a more direct way to increase the matter density of the fireball \citep{2023MNRAS.519.4094W}. However, it also introduces additional physical effects that are not included in the present model. The presence of ions can modify not only the fireball dynamics but also the wave dispersion relation and the linear growth rate of induced scattering. In particular, the linear growth rate in a three component ion-$e^\pm$ plasma has not yet been established. A self-consistent treatment of baryon loading, including its effects on both the outflow dynamics and induced scattering, is therefore left for future work.

\subsubsection{Thermal Evolution and Plasma Heating}
Maintaining the fiducial plasma temperature in the scattering region also requires an additional energy supply. As shown in Appendix~\ref{app:conservative-temperature}, in the absence of additional heating, Compton relaxation and adiabatic expansion can cool the plasma well below the fiducial value $\Theta\sim10^{-2}$ adopted in this work, as given by Eq.~\eqref{eq:conservative-temperature}. According to Eq.~\eqref{eq:frb-kinetic-relation}, the conversion efficiency from kinetic luminosity to FRB luminosity scales approximately as $\Theta^{\frac{1}{2}}$. The conservative temperature estimate can therefore reduce the predicted FRB luminosity by two to three orders of magnitude relative to the fixed temperature case. On the other hand, the energy density of the seed wave can substantially exceed the internal energy density of the particles in the strongly magnetized region considered here. Using Eqs.~\eqref{eq:thermal-energy-density} and \eqref{eq:seed-energy-density}, their ratio at the scattering onset radius in Eq.~\eqref{eq:scattering-onset-radius} is estimated as
\begin{equation}
\begin{aligned}
f_{\mathrm{heat}}
&\sim
\frac{u_{\mathrm{th}}'}{u_{\mathrm{seed}}'}
\sim
\frac{\Theta}
{2\eta_{\mathrm{seed}}'^2\sigma_{\mathrm{B}}'}
\\
&\sim
1.4\times10^{-4}
\left(
\frac{\eta_{\mathrm{seed}}'}{0.5}
\right)^{-\frac{7}{2}}
\left(
\frac{\nu_{\mathrm{obs}}}{1.4\,\mathrm{GHz}}
\right)^{-\frac{3}{4}}
\\
&\quad\times
\frac{
\Gamma_{\mathrm{d},0.5}^{\frac{5}{6}}
\xi_5^{\frac{1}{4}}
\Gamma_{\infty,2.5}^{\frac{5}{4}}
\Theta_{-2}
}{
B_{\mathrm{p},14.3}^{\frac{1}{2}}
R_6
}.
\end{aligned}
\end{equation}
Thus, for the fiducial parameters, transferring only $\sim10^{-4}$ of the seed wave energy to the particle internal energy would be energetically sufficient to maintain $\Theta\sim10^{-2}$. Whether such heating is realized depends on the dissipation mechanism and its efficiency. A self-consistent treatment of wave dissipation and plasma heating is beyond the scope of this work and remains an important subject for future study.

\subsection{Seed Waves and Competing Processes}
\label{sec:seed-waves-competing-processes}
\subsubsection{Origin of the Seed Wave}
For a Lorentz factor of $\Gamma\sim10^{2.5}$ at the scattering onset radius, a seed wave with a frequency of approximately $10\,\mathrm{kHz}$ in the lab frame is required to produce a GHz FRB, as shown in Eq.~\eqref{eq:seed_wave_require}. Possible sources include waves directly excited by starquakes \citep{Blaes1989-starquakes,2026ApJ...998..190Q}, FMS waves produced through wave--wave conversion \citep{2021ApJ...908..176Y}, and small-scale FMS waves generated by magnetic reconnection \citep{Yuan2020-tu,Yuan2022-gk}. As shown in Sec.~\ref{sec:seed-wave-properties}, the seed wave required in our model carries an energy much smaller than those associated with the background dipole magnetic field and the X-ray burst. 

\subsubsection{Competing Induced Scattering Modes}
We focus on induced Compton scattering through the neutral mode. Other branches of induced scattering can also operate under different plasma conditions. Within the neutral mode, stimulated Brillouin scattering can dominate \citep{Nishiura2026-nx}. However, induced Compton scattering occupies a large fraction of the parameter space relevant to strongly magnetized magnetar environments with $\sigma_B\gg1$ \citep{kmdy-17md}. Induced scattering can also excite the charged mode in addition to the neutral mode. We show below that the charged mode is strongly suppressed in the high density regime considered in this work. In the limit $\omega_0'\ll\omega_{\mathrm{p}}'\ll\omega_{\mathrm{c}}$, the maximum linear growth rate of the charged mode is given by Eq.~(125) in \citep{Nishiura2026-nx} as
\begin{equation}
\begin{aligned}
\gamma_{\mathrm{C}}'
&=
16\pi
\eta_{\mathrm{seed}}'^{\,2}
\Theta^{2}
\frac{\omega_0'^{3}}
{\omega_{\mathrm{p}}'^{2}}.
\end{aligned}
\end{equation}
Using Eq.~\eqref{eq:neutral-growth-rate}, the ratio of the charged mode to neutral mode growth rates at the scattering onset radius in Eq.~\eqref{eq:scattering-onset-radius} is estimated as
\begin{equation}
\label{eq:charged_vs_neutral}
\begin{aligned}
\frac{\gamma_{\mathrm{C}}'}
{\gamma_{\mathrm{N}}'}
&=
16\Theta^{2}
\frac{\omega_0'^{2}}
{\omega_{\mathrm{p}}'^{2}}
\sigma_{B}'
\\
&=
1.5\times10^{-10}
\left(
\frac{\nu_{\mathrm{obs}}}
{1.4\,\mathrm{GHz}}
\right)^2
\frac{\Theta_{-2}^{\,2}
B_{\mathrm{p},14.3}^{2}
R_6^{2}
}{
\Gamma_{\mathrm{d},0.5}^{\frac{20}{3}}
\xi_5^{2}
}
\ll 1.
\end{aligned}
\end{equation}
The charged mode therefore makes only a negligible contribution in the high density parameter range considered here.

\subsubsection{Competing Three Wave Interactions}
Three wave interactions between Alfv\'en and FMS waves may also compete with induced scattering \citep{1998PhRvD..57.3219T,2019ApJ...881...13L,2019MNRAS.483.1731L,2023ApJ...957..102G}. However, in the geometry considered here, the force free growth rate of this interaction may be strongly suppressed by the factor $\sin^4\theta_0'$ for a broadband seed wave \citep{2026arXiv260619448S}. An FMS wave incident on the fireball outflow at a lab frame angle $\theta_0=\mathcal{O}(1)$ is aberrated in the comoving frame according to
\begin{equation}
\begin{aligned}
\sin\theta_0'
&=
\frac{
\sin\theta_0
}{
\Gamma_\infty
\left(
1-\beta\cos\theta_0
\right)
}=
\mathcal{O}
\left(
\Gamma_\infty^{-1}
\right)
\ll 1.
\end{aligned}
\end{equation}
The seed wave therefore propagates nearly antiparallel to the local dipole magnetic field in the comoving frame. This geometric suppression suggests that induced scattering through the neutral mode can remain important compared with three wave interactions. A quantitative comparison of the competing processes is left for future work.

\subsubsection{Linearity of the Generated FRB}
The generated FRB also remains a small amplitude wave relative to the background magnetic field at the production radius. The comoving energy density of the FRB is characterized by the saturation energy density of induced scattering in Eq.~\eqref{eq:saturation-energy-density}, so that $u_{\mathrm{FRB}}'\simeq u_{\mathrm{sat}}'\simeq \delta B_{\mathrm{FRB}}'^2/(4\pi)$. Defining the relative magnetic amplitude in the comoving frame as $\eta_{\mathrm{FRB}}'\equiv \delta B_{\mathrm{FRB}}'/B_{\mathrm{d}}$, its value at the scattering onset radius in Eq.~\eqref{eq:scattering-onset-radius} is estimated as
\begin{equation}
\begin{aligned}
\eta_{\mathrm{FRB}}'
&=
\frac{
\sqrt{4\pi u_{\mathrm{sat}}'}
}{
B_{\mathrm{d}}
}
=
\frac{
\Theta^{\frac{1}{4}}
}{
2\sqrt{\sigma_{\mathrm{B}}'}
}
\\
&=
1.3\times10^{-2}\left(
\frac{\eta_{\mathrm{seed}}'}
{0.5}
\right)^{-\frac{3}{4}}
\left(
\frac{\nu_{\mathrm{obs}}}
{1.4\,\mathrm{GHz}}
\right)^{-\frac{3}{8}}\\
&\quad\times
\frac{\Theta_{-2}^{\,\frac{1}{4}}
\Gamma_{\mathrm{d},0.5}^{\frac{5}{12}}\Gamma_{\infty,2.5}^{\frac{5}{8}}
\xi_5^{\frac{1}{8}}
}{
B_{\mathrm{p},14.3}^{\frac{1}{4}}
R_6^{\frac{1}{2}}
}
\ll 1,
\end{aligned}
\end{equation}
which is identical to Eq.~\eqref{eq:saturation-seed-amplitude}, since the generated FRB reaches the same saturation energy density $u_{\mathrm{sat}}'$ that defines the partial--full scattering boundary. Thus, the scattered wave can be treated as a small amplitude perturbation relative to the background magnetic field at the FRB production radius. Propagation at larger radii and the final escape of the generated FRB are beyond the scope of this work and will be investigated in future studies.

\subsection{Lorentz Factor and Viewing Geometry}
\label{sec:lorentz-factor-viewing-geometry}
The predicted FRB luminosity can depend strongly on the terminal Lorentz factor of the outflow. In this work, we adopt $\Gamma_{\infty}=10^{2.5}$ as a fiducial value. In particular, the FRB luminosity at the local freeze out limit has a strong dependence on the Lorentz factor, $L_{\mathrm{FRB}}^{\mathrm{iso}}\propto\Gamma_{\infty}^{19/6}$, as shown in Eq.~\eqref{eq:frb-luminosity-freeze-iso}. However, for a sufficiently large terminal Lorentz factor, induced scattering can become efficient before the outflow reaches the coasting radius in Eq. \eqref{eq:acceleration-radius}. In this case, the FRB luminosity is determined by the Lorentz factor at the scattering radius rather than by $\Gamma_{\infty}$, and the luminosity derived under the coasting approximation cannot be directly extrapolated to larger $\Gamma_{\infty}$. For the local freeze out case, the transition occurs approximately when the scattering onset radius becomes comparable to the coasting radius, corresponding to $\Gamma_{\infty}\sim180$. For larger $\Gamma_{\infty}$, the coasting estimate can therefore overestimate the FRB luminosity.

Rotation of the magnetar can increase the fraction of sources that are geometrically observable over a long period. For simplicity, we assume that the radiation is confined to a single cone with a half opening angle $\theta_{\mathrm{b}}\simeq1/\Gamma\ll1$ and that the beam axis is fixed to the magnetic axis. For a randomly oriented line of sight, the geometrical probability of observing a single burst is given by
\begin{equation}
\label{eq:single-burst-viewing-probability}
\begin{aligned}
P_{\mathrm{burst}}
&=\frac{1}{4\pi}\int_0^{2\pi}\dd\phi\int_0^{\theta_{\mathrm{b}}}\sin\theta\dd\theta
\simeq
\frac{\theta_{\mathrm{b}}^2}{4}
\simeq
\frac{1}{4\Gamma^2}
\\
&\simeq
2.5\times10^{-4}\%
\,\Gamma_{2.5}^{-2}.
\end{aligned}
\end{equation}
In contrast, rotation allows the beam to sweep across a larger region of the sky. Let $\alpha$ be the angle between the magnetic and rotation axes, and let $\zeta$ be the angle between the line of sight and the rotation axis. The line of sight enters the emission cone at some rotational phase when $|\zeta-\alpha|<\theta_{\mathrm{b}}$. For $\theta_{\mathrm{b}}<\alpha<\pi-\theta_{\mathrm{b}}$, the fraction of randomly oriented sources whose swept regions intersect the line of sight is given by
\begin{equation}
\label{eq:rotation-swept-viewing-probability}
\begin{aligned}
P_{\mathrm{sweep}}
&=
\frac{2\pi}{4\pi}
\int_{\alpha-\theta_{\mathrm{b}}}^{\alpha+\theta_{\mathrm{b}}}
\sin\zeta\,\dd\zeta\simeq
\frac{\sin\alpha}{\Gamma}
\\
&
\simeq
0.32\%
\left(\frac{\sin\alpha}{1}\right)
\Gamma_{2.5}^{-1}.
\end{aligned}
\end{equation}
Thus, for $\sin\alpha=\mathcal{O}(1)$, the probability of observing an individual burst scales as $\mathcal{O}(\Gamma^{-2})$, whereas the fraction of sources that are geometrically observable over sufficiently long monitoring scales as $\mathcal{O}(\Gamma^{-1})$. If the magnetic and rotation axes are nearly aligned, the swept region becomes much smaller and this enhancement is reduced. However, for a line of sight that lies within the narrow emission cone, the source can remain visible over a larger fraction of the rotational phase, potentially increasing the observed repetition rate.

\subsection{Observable Properties}
\label{sec:observable-properties}
\subsubsection{Temporal Structure}
\label{sec:observable-timescale}
The relativistic motion of the scattering region introduces a characteristic timescale in our scenario. For example, for a scattering radius of $r_{\mathrm{sc}}\sim10^9\,\mathrm{cm}$ and a Lorentz factor of $\Gamma\sim10^{2.5}$, the relativistic arrival time difference is estimated as
\begin{equation}
\label{eq:timescale}
\begin{aligned}
t_{\mathrm{obs}}
&\sim
\frac{r_{\mathrm{sc}}}{2c\Gamma^2}
\simeq
0.2\,\mu{\rm s}~
\frac{r_{\mathrm{sc},9}}
{\Gamma_{2.5}^{2}}.
\end{aligned}
\end{equation}
This timescale is consistent with the microsecond and submicrosecond structures observed in some FRBs \citep{Nimmo2021-microstructure,Majid2021-nanostructure,Snelders2023-ultrafast}. For the small pair loading region discussed in Sec.~\ref{sec:small}, with $\Delta\theta_B\ll1/\Gamma$, the corresponding arrival time difference can be even shorter. Eq.~\eqref{eq:timescale}, however, does not necessarily determine the overall FRB duration. For example, if the central engine sustains a continuous fireball outflow for $\Delta t_{\mathrm{flow}}\gg t_{\mathrm{obs}}$, the observed burst duration may instead be characterized by the outflow duration.

\subsubsection{Frequency Range}
\label{sec:observable-frequency}
The characteristic frequency of the generated FRB is not restricted to the GHz band in this mechanism. As shown in Eq.~\eqref{eq:seed_wave_require}, the observed frequency is determined by the combination of the seed wave frequency and the Lorentz factor at the scattering radius. The same scattering process can therefore produce radiation at frequencies different from the GHz band for different seed wave frequencies or outflow Lorentz factors.

\subsubsection{Polarization}
\label{sec:observable-polarization}
For a linearly polarized FMS seed wave, induced scattering preferentially amplifies a corresponding linear polarization state. The neutral mode growth rate is maximized when the electric fields of the seed and scattered waves are parallel \citep{2025PhRvD.111f3055N,Nishiura2026-nx}. A high degree of linear polarization can therefore be produced when the seed wave is strongly linearly polarized. More generally, even a factor of order unity difference between the growth rates of different polarization states can generate a large contrast after exponential amplification. Induced scattering can therefore produce strong polarization selectivity.

The present scenario can also produce circularly polarized FRBs, as demonstrated in one dimensional PIC simulations \citep{tvyv-yn1z,kmdy-17md}. For a circularly polarized Alfv\'en incident wave propagating parallel to the background magnetic field, circularly polarized backward scattered waves grow through both the charged and neutral modes \citep{tvyv-yn1z,kmdy-17md}. These results show that strong circular polarization can be produced from the seed wave to the scattered radiation. Whether this property is preserved in multidimensional geometries with oblique scattering remains to be investigated. Such polarization selectivity can provide a possible explanation for the circular polarization fractions approaching $90\%$ observed in some FRBs \citep{2024NSRev..12E.293J}.

More complex polarization behavior may arise if the dominant induced scattering mode changes between the neutral and charged modes within the emission region. Such orthogonal polarization angle jumps have been observed in some FRBs \citep{2024ApJ...972L..20N}. In the comoving frame, for propagation parallel or antiparallel to the background magnetic field, the fastest growing scattered wave in the neutral mode has an electric field parallel or antiparallel to that of the seed wave. In contrast, the charged mode preferentially amplifies a scattered wave whose electric field is orthogonal to that of the seed wave \citep{2025PhRvD.111f3055N,Nishiura2026-nx}. A transition in the dominant mode can therefore produce a $90^\circ$ change in the observed polarization angle. For the fiducial high density parameters adopted in this work, the charged mode is strongly suppressed, as shown in Eq.~\eqref{eq:charged_vs_neutral}, and such a transition is not expected. This effect may nevertheless become important in scattering regions with different plasma conditions.

\section{Conclusion}
\label{sec:conclusion}
We have developed a model in which FRBs are produced by induced Compton upscattering in an $e^\pm$ fireball outflow formed near the magnetic pole of a magnetar. Our results identify the plasma conditions under which induced Compton upscattering can connect magnetar X-ray activity to coherent radio bursts. Escaping X-rays accelerate the $e^\pm$ plasma to a relativistic bulk Lorentz factor. Low frequency FMS or Alfv\'en waves with frequencies of order $10\,\mathrm{kHz}$ can then be upscattered into the GHz band when they enter the outflow from the side or against its bulk motion. We estimate the luminosity of the scattered radiation by combining the dynamics of the fireball outflow, the linear growth rate of induced scattering, and a prescription for nonlinear saturation motivated by PIC simulations.

The FRB luminosity is primarily controlled by the kinetic luminosity and comoving temperature of the fireball outflow and is approximately expressed as
$L_{\rm FRB}^{\rm iso}\sim\Theta^{\frac{1}{2}}L_{\rm kin}^{\rm iso}$.
For the fireball parameters inferred from the X-ray burst associated with Galactic FRB 20200428, the conservative post-freeze-out pair density alone is insufficient to reproduce the observed radio luminosity, indicating that additional pair loading is required.
However, with enhanced pair loading, the model can account for the luminosity of Galactic FRB 20200428. At still higher densities approaching the local freeze-out limit, the model can reach luminosities comparable to those of the brightest extragalactic FRBs.
Thus, the pair density and plasma temperature are the key physical quantities that determine the luminosity range achievable in this scenario.

We have also examined the spectral bandwidth produced by a localized pair loading region whose angular extent is smaller than the relativistic beaming angle. For viewing directions outside the velocity region but within the relativistic beaming cone, geometrical Doppler broadening gives
$L_{\rm FRB}^{\rm iso}\propto(\Delta\nu/\nu_{\rm c})^2$.
The observed burst distributions of the three repeating FRBs considered in this work are qualitatively consistent with this trend. A quantitative test will require a more detailed treatment of observational selection effects and the truncation of burst spectra at the edges of the observing bands.

Several important issues remain open. In particular, the supply and heating of the $e^\pm$ plasma should be determined self-consistently, and the propagation and escape of the generated radio waves require further investigation. This work has focused on the polar fireball channel. In a subsequent paper, we will investigate an off-polar channel in which a magnetically accelerated plasmoid fireball forms near the equatorial region of the magnetar.

\begin{acknowledgments}
We gratefully acknowledge insightful discussions with Kazumi Kashiyama and Tomoki Wada. RN is supported by JSPS KAKENHI, Grant No. 25KJ1562. KI is supported by MEXT/JSPS KAKENHI Grant No. 26H02045, 23H01172, 23H05430, 23H04900, 22H00130. 
\end{acknowledgments}
\appendix
\section{Full Scattering Regime of Induced Scattering}
\label{app:full-scattering}
For completeness, we consider the full scattering regime, in which
$u_{\rm seed}'<u_{\rm sat}'$. In this case, the seed wave is
substantially attenuated before the scattered wave reaches the
saturation energy density in Eq. \eqref{eq:saturation-energy-density}. The scattered wave energy density is
therefore limited by the seed wave energy density, and the resulting
energy flux is estimated as
\begin{equation}
\label{eq:full-scattering-flux}
\begin{aligned}
F_{\rm full}
&\leq cu_{\rm seed}
\\
&\sim c\Gamma^2
\left\langle
\left(1+\beta\cos\theta_1'\right)^2
\right\rangle_E
u_{\rm seed}'
\\
&\sim c\Gamma^4
\frac{\delta B_{\rm seed}^2}{4\pi}.
\end{aligned}
\end{equation}
Combining this limit with the partial scattering flux in
Eq.~\eqref{eq:saturation-flux}, the FRB isotropic luminosity can be
written generally as
\begin{equation}
\label{eq:frb-luminosity-general}
\begin{aligned}
L_{\rm FRB}^{\rm iso}
&\sim 4\pi r^2
\min\left\{F_{\rm sat},\,F_{\rm full}\right\}
\\
&\sim \min\left\{
\Theta^{\frac{1}{2}}L_{\rm kin}^{\rm iso},
\Gamma^4L_{\rm seed}^{\rm iso}
\right\}.
\end{aligned}
\end{equation}
Thus, the FRB luminosity is limited by the kinetic luminosity of the
fireball outflow in the partial scattering regime and by the seed wave
luminosity in the full scattering regime. The calculations in the main
text focus on the partial scattering branch relevant to the parameter
range considered in this work.

\section{Conservative Estimate of the Comoving Temperature}
\label{app:conservative-temperature}
We estimate the comoving plasma temperature when no additional dissipative heating operates after photon decoupling. Although the X-ray radiation is no longer trapped in the $e^\pm$ plasma, the escaping photons can continue to exchange energy with the pairs through Compton scattering. As long as Compton relaxation is faster than the expansion, this interaction drives the plasma temperature toward the Compton temperature of the radiation field \citep{1979rpa..book.....R}. We approximate the X-ray photon number spectrum per unit volume and unit energy by a cutoff power law,
\begin{equation}
\frac{\dd N_\gamma}{\dd E}\propto E^{-\Gamma_\gamma}
\exp\left(
-\frac{E}{E_{\mathrm{cut}}}
\right).
\end{equation}
In the Thomson limit, the Compton temperature is obtained by
setting the spectrum averaged energy exchange between photons and
electrons to zero \citep{1979rpa..book.....R}. This gives
\begin{equation}
\label{eq:Compton_cutoff}
\begin{aligned}
k_{\mathrm{B}}T_{\mathrm{C}}
&=
\frac{1}{4}
\frac{
\displaystyle \int_0^\infty
E^2 \frac{dN_\gamma}{dE}\,dE
}{
\displaystyle \int_0^\infty
E \frac{dN_\gamma}{dE}\,dE
}
\\
&=
\frac{1}{4}
\frac{
\Gamma(3-\Gamma_\gamma)
}{
\Gamma(2-\Gamma_\gamma)
}
E_{\mathrm{cut}}
=
\frac{2-\Gamma_\gamma}{4}
E_{\mathrm{cut}},
\end{aligned}
\end{equation}
where $\Gamma(\cdots)$ denotes the gamma function.

The X-ray spectrum associated with Galactic FRB 20200428 implies a Compton temperature of order $10\,\mathrm{keV}$. Fitting the Insight-HXMT spectrum with a cutoff power law, \citet{Li2021-nd} obtained
\begin{equation}
\label{eq:Compton_observation}
\begin{aligned}
\Gamma_\gamma
&\simeq
1.56\pm0.06,
\\
E_{\mathrm{cut}}
&=
83.89^{+9.08}_{-7.55}\ {\rm keV}.
\end{aligned}
\end{equation}
Substituting Eq.~\eqref{eq:Compton_observation} into Eq.~\eqref{eq:Compton_cutoff} gives
\begin{equation}
\begin{aligned}
k_{\mathrm{B}}T_{\mathrm{C}}
\sim
10\ {\rm keV}.
\end{aligned}
\end{equation}
We therefore assume that, after photon decoupling, Compton scattering maintains the plasma near this temperature as long as the Compton relaxation time remains shorter than the comoving dynamical time. This approximation ceases to hold beyond the radius $r_{\mathrm{C}}$ defined below in Eq.~\eqref{eq:compton-decoupling-radius}.

We next determine the thermal evolution after the plasma begins to decouple thermally from the X-ray radiation. We approximate the escaping X-ray radiation as a radial beam propagating outward from behind the plasma and adopt an adiabatic index of $\gamma_{\mathrm{ad}}=5/3$ for $e^\pm$ plasma.

The Compton relaxation time for nonrelativistic particles at rest in a radiation field with energy density $u_{\mathrm{X}}$ is given by
\begin{equation}
\label{eq:compton-cooling-time}
t_{\mathrm{C}}
=
\frac{3m_{\mathrm{e}}c^2}
{8\sigma_{\mathrm{T}}cu_{\mathrm{X}}}.
\end{equation}
For a radial photon beam propagating in the same direction as the relativistic outflow, the radiation is deboosted in the comoving frame. Its comoving energy density is approximated as
\begin{equation}
\label{eq:comoving-xray-energy-density}
u_{\mathrm{X}}'(r)
\simeq
\frac{u_{\mathrm{X}}(r)}{4\Gamma(r)^2}
=
\frac{L_{\mathrm{X}}}{16\pi r^2c\Gamma(r)^2}.
\end{equation}
Transforming Eq.~\eqref{eq:compton-cooling-time} to the comoving frame and using Eqs.~\eqref{eq:comoving-xray-energy-density}, \eqref{eq:gamma-decoupling-scaling}, and \eqref{eq:post-freeze-lorentz-factor}, the relaxation time is expressed as
\begin{equation}
\label{eq:compton-cooling-time-radius}
\begin{aligned}
t_{\mathrm{C}}'(r)
&=
\frac{3m_{\mathrm{e}}c}{8\sigma_{\mathrm{T}}u_{\mathrm{X}}'(r)}
\\
&\simeq
5.0\times10^{-6}\,\mathrm{s}\,
\frac{r_7^4\Gamma_{\mathrm{d},0.5}^{\frac{2}{3}}}
{L_{\mathrm{X},41}R_6^2}.
\end{aligned}
\end{equation}

We define the Compton decoupling radius $r_{\mathrm{C}}$ by equating the Compton relaxation time with the comoving dynamical time. During the acceleration phase, this condition is given by
\begin{equation}
\label{eq:compton-decoupling-condition}
\begin{aligned}
t_{\mathrm{C}}'(r_{\mathrm{C}})
&=t_{\mathrm{dyn}}'(r_{\mathrm{C}})
=\frac{R}{c\Gamma_{\mathrm{d}}^{\frac{1}{3}}}
\\
&\simeq
2.3\times10^{-5}\,\mathrm{s}\,
R_6\Gamma_{\mathrm{d},0.5}^{-\frac{1}{3}}.
\end{aligned}
\end{equation}
The resulting Compton decoupling radius is
\begin{equation}
\label{eq:compton-decoupling-radius}
\begin{aligned}
r_{\mathrm{C}}
&=
\left(
\frac{\sigma_{\mathrm{T}}L_{\mathrm{X}}R^3}
{6\pi m_{\mathrm{e}}c^3\Gamma_{\mathrm{d}}}
\right)^{\frac{1}{4}}
\\
&\simeq
1.5\times10^7\,\mathrm{cm}\,
\frac{L_{\mathrm{X},41}^{\frac{1}{4}}R_6^{\frac{3}{4}}}
{\Gamma_{\mathrm{d},0.5}^{\frac{1}{4}}}.
\end{aligned}
\end{equation}
For the fiducial parameters, Eq.~\eqref{eq:acceleration-radius} gives
$r_{\mathrm{acc}}=2.2\times10^8\,\mathrm{cm}$, and hence
$r_{\mathrm{C}}/r_{\mathrm{acc}}=0.07$. The assumed ordering
$r_{\mathrm{C}}<r_{\mathrm{acc}}$ is therefore well satisfied.

While Compton relaxation remains efficient, the comoving plasma temperature follows the Compton temperature of the deboosted X-ray radiation. For a radial photon beam propagating from behind, the characteristic photon energy in the comoving frame is smaller than that in the lab frame by approximately $2\Gamma$. The comoving temperature at $r_{\mathrm{C}}$ is therefore estimated as
\begin{equation}
\label{eq:temperature-at-compton-decoupling}
\begin{aligned}
\Theta_{\mathrm{C}}(r_{\mathrm{C}})
&\simeq
\frac{k_{\mathrm{B}}T_{\mathrm{C}}}{2\Gamma(r_{\mathrm{C}})m_{\mathrm{e}}c^2}\\
&\simeq
4.6\times10^{-4}
\frac{T_{\mathrm{C},8}R_6^{\frac{1}{4}}}
{L_{\mathrm{X},41}^{\frac{1}{4}}
\Gamma_{\mathrm{d},0.5}^{\frac{1}{12}}}.
\end{aligned}
\end{equation}

Beyond $r_{\mathrm{C}}$, Compton energy exchange becomes inefficient and we approximate the subsequent thermal evolution as adiabatic. The ideal gas relations are
\begin{equation}
\label{eq:adiabatic-cooling-relations}
\left\{
\begin{aligned}
&P'V'^{\gamma_{\mathrm{ad}}}=\mathrm{const},
\\
&P'=n_\pm'k_{\mathrm{B}}T',
\\
&n_\pm'V'=\mathrm{const}.
\end{aligned}
\right.
\end{equation}
For $\gamma_{\mathrm{ad}}=5/3$, these relations give
$\Theta\propto(n_\pm')^{2/3}$.
Particle number flux conservation in a dipolar flux tube gives
$n_\pm'\propto(\Gamma r^3)^{-1}$.
During radiative acceleration, $\Gamma\propto r$, so that
$n_\pm'\propto r^{-4}$ and $\Theta\propto r^{-8/3}$.
The temperature at the end of the acceleration phase is therefore expressed as
\begin{equation}
\label{eq:temperature-at-acceleration-radius}
\begin{aligned}
\Theta(r_{\mathrm{acc}})
&\simeq
\Theta_{\mathrm{C}}(r_{\mathrm{C}})
\left(\frac{r_{\mathrm{C}}}{r_{\mathrm{acc}}}\right)^{\frac{8}{3}}
\\
&\simeq
3.5\times10^{-7}
\frac{T_{\mathrm{C},8}
L_{\mathrm{X},41}^{\frac{5}{12}}
\Gamma_{\mathrm{d},0.5}^{\frac{5}{36}}
}{
R_6^{\frac{5}{12}}
\Gamma_{\infty,2.5}^{\frac{8}{3}}
}.
\end{aligned}
\end{equation}

After radiative acceleration ends, the outflow enters the coasting regime with approximately constant $\Gamma$. Particle number flux conservation then gives $n_\pm'\propto r^{-3}$, and adiabatic cooling gives $\Theta\propto r^{-2}$. The conservative temperature at the scattering onset radius in Eq.~\eqref{eq:scattering-onset-radius} is therefore given by
\begin{equation*}
\label{eq:temperature-at-scattering-radius-conservative}
\begin{aligned}
\Theta_{\mathrm{con}}(r_{\mathrm{sc}})
&\simeq
\Theta'(r_{\mathrm{acc}})
\left(\frac{r_{\mathrm{acc}}}{r_{\mathrm{sc}}}\right)^2
\\
&\simeq
4.3\times10^{-8}
\left(\frac{\eta_{\mathrm{seed}}'}{0.5}\right)
\left(\frac{\nu_{\mathrm{obs}}}{1.4\,\mathrm{GHz}}\right)^{\frac{1}{2}}
\\
&\quad\times
\frac{T_{\mathrm{C},8}
L_{\mathrm{X},41}^{\frac{5}{12}}
\Gamma_{\mathrm{d},0.5}^{\frac{41}{36}}
\xi_5^{\frac{1}{2}}
}{
R_6^{\frac{5}{12}}
\Gamma_{\infty,2.5}^{\frac{13}{6}}
B_{\mathrm{p},14.3}
},
\end{aligned}
\end{equation*}
which recovers Eq.~\eqref{eq:conservative-temperature}.

\nocite{*}
\bibliographystyle{apsrev4-2}
\bibliography{apssamp}
\end{document}